\documentclass[aps,prl,reprint,nobalancelastpage,nofootinbib]{revtex4-2}

\usepackage[utf8]{inputenc}

\usepackage{amsmath}
\usepackage{amssymb}
\usepackage[all]{xy}
\usepackage{tikz}
\usetikzlibrary{calc}

\usepackage[toc,page]{appendix}
\usepackage{enumerate}
\usepackage{tensor}
\usepackage{booktabs}
\usepackage{bbm}
\usepackage{youngtab}
\usepackage{slashed}
\usepackage{multirow}
\usepackage{makecell}
\usepackage{float}
\usepackage{comment}
\usepackage{verbatim}
\usepackage{xcolor}
\usepackage[framemethod=tikz]{mdframed}
\usepackage{soul}
\usepackage{mathtools}
\usepackage[inline]{enumitem}
\usepackage{alphabeta}

\usepackage{amsthm}

\newcommand{\ab}{|}
\newcommand{\der}{\partial}
\newcommand{\de}{\mathrm{d}}

\newcommand{\e}{\mathrm{e}}
\newcommand{\I}{\mathrm{i}}

\usepackage{hyperref}
\hypersetup{
    colorlinks=true,
    linkcolor=cyan,
    citecolor=cyan,
    filecolor=cyan,      
    urlcolor=cyan,
}

\newcommand{\MMVV}{{\mathrm{MMVV}}}
\newcommand{\E}{{\mathrm{E}}}

\makeatletter
\renewcommand{\p@subsection}{}
\renewcommand{\p@subsubsection}{}
\makeatother

\begin{document}
\numberwithin{equation}{section}

\title{A distance conjecture beyond moduli?}
\author{Cédric Debusschere}
\email{cedric.debusschere@student.kuleuven.be}
\author{Flavio Tonioni}
\email{flavio.tonioni@kuleuven.be}
\author{Thomas Van Riet}
\email{thomas.vanriet@kuleuven.be}
\affiliation{Instituut voor Theoretische Fysica, KU Leuven, Celestijnenlaan 200D, B-3001 Leuven, Belgium}

\begin{abstract}
The distance conjecture states that for theories with moduli coupled to gravity a tower of states becomes exponentially light in the geodesic distance in moduli space. This specifies how effective field theories break down for large field values. However, phenomenological field theories have no moduli, but a scalar potential that deforms dynamical trajectories away from geodesic curves. In this note, we speculate on how one should generalise the distance conjecture, in asymptotic field regimes, to include a scalar potential. We test the generalized distance conjecture in a few cases, demonstrate a link with pseudo-/fake supersymmetry and apply it to the ekpyrotic scenario in cosmology. For the latter we observe that the pre-uplift KKLT potential could provide a stringy embedding of ekpyrosis away from asymptotic regimes in field space.
\end{abstract}

\maketitle

\section{Introduction}

Effective field theories (EFTs) can be thought of as being described by Lagrangians for which the relevant part is typically the renormalisable sector and the irrelevant terms group into a series expansion in higher derivatives and higher field powers, suppressed by the cut-off scale $\Lambda_C$.
The EFT breaks down when field values or gradients become as big as $\Lambda_C$ since one would need to know an ever increasing number of Wilson coefficients, for a probably non-convergent series expansion anyway. 

In the presence of a ultraviolet (UV) completion of gravity, it has been possible to quantify how large field values invalidate an EFT through the so-called \emph{distance conjecture} \cite{Ooguri:2006in}.
For the latter, vast evidence in top-down string theory compactifications and supportive arguments beyond circumstantial evidence have been gathered \cite{Stout:2021ubb}; see e.g. refs.~\cite{Palti:2019pca, vanBeest:2021lhn, Agmon:2022thq} for reviews.

As one travels far in moduli space, a tower of degrees of freedom (states) originating from the UV completion becomes light and changes the EFT.
Typical examples are the string vibration modes or Kaluza-Klein (KK) particles \cite{Lee:2019xtm, Lee:2019wij}.
The distance conjecture quantifies this breakdown as follows.
Let $m$ be the tower mass gap and $\Delta$ the geodesic distance travelled in moduli space, starting from a reference point where the EFT is under control, and thus $m > \Lambda_C$.
The conjecture states that
\begin{equation}\label{distance}
m(\Delta) = m(0) \, \e^{-\alpha \Delta}\,,
\end{equation}
where $\alpha$ is an order-1 number and $\Delta$ is measured in Planck units.
In other words, the EFT breaks down because fields in the UV become lighter such that, at some distance, one finds $m < \Lambda_C$.
The distance conjecture is an integral part of the most well-established Swampland conjectures \cite{Vafa:2005ui} and it applies to pure moduli spaces, which can exist in supersymmetric (SUSY) theories.

What seems to be lacking is a similar statement for large field gradients, which imply a breakdown also from the EFT viewpoint.
How does one incorporate this in the quantum gravity (Swampland) context?
If one considers just time evolution, then field gradients are measured with time-derivatives of the fields, and thus the momenta.
Without a scalar potential, momenta projected along the geodesic are covariantly conserved. A scalar potential would ruin this, which suggests a distance formula might exist in momentum space as well, where large momentum variations lead to a breakdown of the EFT by integrating in new modes.
Given that Hamiltonian dynamics is symplectic-covariant in phase space, and that flipping momenta with coordinates $(q^i,p_i)\leftrightarrow (p_i,-q^i)$ is a canonical transformation, one might be led to hypothesize that distances in \emph{phase space} are fundamental.
Formulating a distance conjecture in phase space therefore seems like a natural idea, but it appears to be obstructed for simple reasons, which we comment on below.
Instead, we argue for (and test) an extension of the distance conjecture where the distance is measured by integrating the square root of twice the total field energy (which is not constant in presence of gravity) along an on-shell trajectory $\mathrm{C}$, i.e.
\begin{equation} \label{generalized distance}
\Delta = \int_{\mathrm{C}} \de \tau \, \sqrt{2T + 2V}\,,
\end{equation}
where $T$ is the kinetic energy and $V$ is the potential.
We refrain from computing distances inside the field-space bulk, and we propose that the on-shell trajectory $\mathrm{C}$ and the coordinate dependence should be taken along the attractor solution to the field equations, coupled to gravity.
This definition allows one to consistently move from an initial point towards the field-space asymptotics, in the spirit of the original distance conjecture, and is independent of the initial conditions.\footnote{We consider attractors that are defined to flow towards the field-space asymptotics, independently of initial conditions. Hence, our proposal to compute distances from points in the asymptotics is also independent of the initial conditions.}
When $V=0$, our definition collapses to that of the ordinary distance conjecture. 
Different notions of a generalized distance were proposed earlier in refs.~\cite{Schimmrigk:2018gch, Basile:2023rvm} (see also refs.~\cite{Basile:2022zee, Basile:2022sda}) and very recently in ref.~\cite{Mohseni:2024njl}. 

The ordinary distance conjecture has been used to argue why, in asymptotic regimes of field space, positive scalar potentials are not tunably flat, and why local de Sitter (dS) minima are forbidden \cite{Ooguri:2018wrx, Hebecker:2018vxz}.
One argument is that $\Lambda_C$ goes down exponentially with the large field distance while one demands that the Hubble density $H$ be $H<\Lambda_C$ \cite{Hebecker:2018vxz}.
Alternatively, one finds that the entropy counted by $H$ should at least comprise the entropy associated to the light degrees of freedom \cite{Ooguri:2018wrx}.
This is a great success of the distance conjecture, so any generalisation that includes $V$ in the notion of distance should not ruin this success.
We will find that this is indeed the case.
At the same time, though, we want an explanation as to why in the asymptotic regime (10d supergravity compactifications) there neither exist the opposite of shallow positive potentials, i.e. steep negative potentials \cite{Uzawa:2018sal}.
We do not find an answer to this by using the generalized distance conjecture, but we find constraints for ekpyrotic cosmological models which rely on steep negative potentials.

\section{Flows and generalized distances}

After choosing units such that $\kappa_D=1$, we consider the following theory in $D=1+d$ dimensions:
\begin{equation}
S = \int \star_D \; \bigl[\tfrac{1}{2} \mathcal{R}_D - \tfrac{1}{2} G_{ij}(\phi) \partial \phi^i \partial \phi^j - V(\phi) \bigr]\,.
\end{equation}
As dynamical solutions, we consider both spatially-flat Friedmann-Lemaître-Robertson-Walker (FLRW) cosmologies and static domain-wall (DW) type metrics with Minkowski slicing.
This can be captured within one notation as
\begin{equation} \label{metric}
    ds^2_{d+1} = \epsilon N^2(\tau) \de \tau^2 + a^2(\tau) \, (\eta_{\epsilon})_{ab} \, \de x^a \de x^b\,,
\end{equation}
where $\epsilon=-1$ corresponds to FLRW cosmologies, with $\eta_{-1}$ the Euclidean flat metric on the spatial slice, and $\epsilon=+1$ corresponds to Minkowski-sliced domain walls, with $\eta_{+1}$ the Minkowski metric. Such spacetimes require the scalars to only depend on $\tau$, which is a time coordinate if $\epsilon=-1$ and a spatial coordinate if $\epsilon=+1$.
The equations of motion can then be derived
from a 1d classical action for the scalars $\phi^i$, the metric function $a$ and the Lagrange multiplier $N$.
Throughout this note, we employ the notation $\dot{} = \de / \de \tau$.

The reasons why we focus on backgrounds of the form in eq. (\ref{metric}) are simple.
In the absence of a potential, field-space geodesics emerge in the form of FLRW- or DW-type solutions.
More precisely, by restricting the metric to FLRW/DW solutions, the scalar profiles become geodesics with the harmonic function on said spacetimes as an affine coordinate \cite{Bergshoeff:2008be}.
Moreover, in the absence of a potential, the FLRW distance equals the DW distance.
Hence, in the case of non-vanishing potentials, the same class of metrics is a very reasonable ansatz to postulate.
As a secondary reason, cosmological and domain-wall solutions are expected to be simple universal solutions of generic string compactifications in asymptotic regimes.
Exponential potentials for canonical fields are expected to appear in the field-space asymptotics \cite{Ooguri:2018wrx, Hebecker:2018vxz}.
In such cases, the asymptotic attractors are known and therefore make this class of theories ideal candidates for a broad study.
We review and explain each one of these points below.

\subsection{No potential}
Let us consider a theory with $V=0$.
Then, the equations of motion for the scalars can be derived from the 1d effective action
\begin{equation}
S = \int \de \tau \, \frac{a^{D-1}}{2N} \, G_{ij}\dot{\phi}^i\dot{\phi}^j.    
\end{equation}
In the gauge $N=a^{D-1}$, this is the geodesic action and $\tau$ is an affine parameter. The total geodesic velocity $v$ is therefore constant, i.e.
\begin{equation}
    v^2 =  G_{ij}\dot{\phi}^i\dot{\phi}^j = \text{constant},
\end{equation}
and the distance in field space $\Delta$ travelled after a time $\tau$ equals $\Delta = v\tau$.

Let us first consider a single scalar field $\phi$ and its momentum $p=\dot{\phi}$. The phase space is just the flat Euclidean space $\mathbb{E}^2$. In the absence of a potential, the geodesic motion is at constant $p$ and thus the distance travelled in phase space is the distance travelled in configuration space. The distance conjecture extended to phase space trivially works.

What about more multiple fields with non-Riemann flat target spaces?  The geodesic equation of motion in momentum space is
\begin{equation}
p^i \nabla_i p_j =0\,,
\end{equation}
where $p_i=G_{ij}\dot{\phi}^j$, so momenta are covariantly conserved along the geodesic flow. In this sense, there is no motion in momentum space, but one needs this generalized notion of being constant. We can only conclude that $\de(p^ip_i)/\de \tau=0$ and this is a first indication that phase-space distances might not be the right way forward. We have verified that attempts to do this seem to fail. Instead, below we motivate a simple formula that extends the distance conjecture to include for a potential.

\subsection{Motivation: a potential from geodesic motion}
In some circumstances, kinetic energy terms can be seen as potential energy in an effective way. This is for instance well-known in Kepler's problem, where angular momentum acts effectively as a repulsive $1/r^3$-type force in the effective 1d particle motion for the radial evolution. So the angular kinetic particle energy is traded for a $1/r^2$-type potential energy.
If one were to measure the travelled distance, one could simply include this potential. How, exactly? Consider the following kinetic term:
\begin{equation}
L_{\mathrm{kin}} = \tfrac{1}{2} \dot{\phi}^2 + \tfrac{1}{2}f(\phi)\dot{\chi}^2\,.
\end{equation}
The shift symmetry of $\chi$ implies conservation of $\chi$-momentum $Q = f(\phi)\dot{\chi}$. The equation of motion for $\phi$ can be derived from the Lagrangian $L =\dot{\phi}^2/2 - V(\phi)$, where
\begin{equation}
V(\phi) =  \tfrac{1}{2}Q^2 f(\phi)^{-1}\,.    
\end{equation}
Note that the Lagrangian is \emph{not} obtained replacing $\dot{\chi}$ with $Qf^{-1}$ in the original Lagrangian. That would give $V$ with the wrong sign. In other words, the total energy $T+V$, equals the kinetic energy of the original Lagrangian, which measures the distance.

\subsection{The generalized distance (conjecture)}
One can think of any EFT Lagrangian from a compactification as a 10d (or 11d) kinetic term where fields have been integrated out to generate a scalar potential. This is for instance how flux potentials come about. Hence, we propose the following formula for a generalized distance conjecture of the FLRW type:
\begin{equation}\label{masterequation}
\Delta = \int_{\mathrm{C}} N \de \tau \, \sqrt{\dfrac{1}{N^2} \, G_{ij} \dot{\phi}^i\dot{\phi}^j +2V} \,.
\end{equation}
Without gravity, and in the gauge $N=1$, this would simply be $\Delta = \sqrt{2 E} \Delta \tau$, with $E$ the energy. In the presence of gravity, though, the energy $E$ in the field is not conserved.
As we mentioned, scalar potentials appearing in string compactifications induce specific asymptotic behaviors in the form of attractor solutions.
Such asymptotic behaviors are relevant for our discussion since we are interested in probing the field-space boundary.
Hence, we always integrate the total energy along the attractor solution leading towards such asymptotics.
We will discuss these attractor solutions extensively below.

Note that when $V=\Lambda$ we expect the ordinary distance conjecture to hold. So, we make the assumption that $V$ in the above formulas is the potential energy from which the overall vacuum energy $V_0$ in the EFT expansion point\footnote{By ``EFT expansion point'' we mean the vacuum around which the fluctuations group into an EFT, i.e. the point where the potential formally vanishes.} is subtracted:
\begin{equation}
V \rightarrow V - V_0\,.    
\end{equation}
For flows inside a moduli space of e.g. anti-dS vacua, $V$ in the formula should be set to zero.
For exponential potentials that die off to zero at infinity this means $V$ is simply the scalar potential.

What if the evolution is not time-dependent, but spatial? The kinetic term flips sign depending on whether the evolution is spatial or timelike. This simply leads to replacing $V$ in eq.~\eqref{masterequation} with $-\epsilon V$.

\subsection{Link with Hamilton-Jacobi theory}
The proposed extension of the distance conjecture can be captured using the concept of true, fake or pseudo-superpotential \cite{Freedman:2003ax, Skenderis:2006jq, Skenderis:2006rr, deBoer:1999tgo, DiazDorronsoro:2016rrz, Chemissany:2007fg}, as we now explain. 

When the potential in $D$ dimensions can be written in terms of a (fake) superpotential $\tilde{W}$ (for $\epsilon=+1$) or a pseudo-superpotential $\tilde{W}$ (for $\epsilon=-1$) as
\begin{equation}\label{W}
V = \dfrac{\epsilon}{2} \left[ \partial_i \tilde{W} \partial^i \tilde{W} - \frac{D-1}{D-2} \, \tilde{W}^2 \right]\,,
\end{equation}
then, up to total derivatives, the effective action takes a squared form $S = \epsilon\int \de \tau \, a^{D-1} N I$, with
\begin{equation}
\begin{split}
    I = - \dfrac{1}{2} \dfrac{D-1}{D-2} \biggl[ \tilde{W} - (D-2) \frac{\dot{a}}{aN} \biggr]^{2}
    + \dfrac{1}{2} \, \biggl( \dfrac{\dot{\phi}}{N} + \der \tilde{W} \biggr)^{\!\!2},
\end{split}
\end{equation}
where the field-space indices are contracted through the field-space metric, leading to the first-order equations
\begin{subequations}
    \begin{align}
        \dfrac{\dot{\phi}^i}{N} & = - \der^i \tilde{W}, \label{flow 1} \\
        \frac{\dot{a}}{aN} & = \dfrac{\tilde{W}}{D-2}. \label{flow 2}
    \end{align}
\end{subequations}
In the static case, these equations can correspond to actual SUSY conditions, if the superpotential is not fake.
In the cosmological case this is not possible, unless one considers non-unitary supergravity theories.
Regardless of SUSY, if one finds a function $\tilde{W}$ obeying eq.~\eqref{W}, one can solve first-order equations.
This is nothing but a consequence of the Hamilton-Jacobi theory for a principal function $\mathcal{S}(\phi, N, a)$ that factorises as $\mathcal{S} = \tilde{W}(\phi) f(N,a)$ with $f$ some function \cite{DiazDorronsoro:2016rrz}.
So first-order equations should not be a surprise, since they are always (locally) possible, but possibly obstructed at certain points in configuration space, as guaranteed by theorems in symplectic geometry \cite{DiazDorronsoro:2016rrz}.
However, the factorisation property that leads to eq.~\eqref{W} is not obvious.\footnote{One can show that the factorisation can always be done, but then the integration constants in the function $\tilde{W}$ can fix initial conditions for scalars such that the flow equations (\ref{flow 1}, \ref{flow 2}) cannot be seen as a defining a congruence of curves on the scalar manifold defined through the vector field $G^{ij} \partial_j \tilde{W}$.}
In the static case, these flow equations can sometimes be interpreted as renormalization-group equations \cite{deBoer:1999tgo} with $\tilde{W}$ playing the role of the generalized C-function.

In terms of the function $\tilde{W}$, in the gauge $N=1$, the distance travelled in configuration space is
\begin{equation}
\Delta = \int_{\mathrm{C}} \de \tau \, \sqrt{G^{ij}\partial_i \tilde{W}\partial_j \tilde{W}}\,.
\end{equation}
If we use eq.~\eqref{W}, we can replace $G^{ij} \partial_i \tilde{W} \partial_j \tilde{W}$ by $ 2\epsilon V + [(D-1)/(D-2)] \, \tilde{W}^2$. Hence, when $V=0$, we have
\begin{equation} \label{generalized distance - superpotential formulation}
\Delta = \sqrt{\frac{D-1}{D-2}} \int_{\mathrm{C}} \de \tau \, \ab \tilde{W} \ab \,.
\end{equation}
Our suggestion is to use this equation, even when $V\neq 0$. This amounts to measuring distances exactly according to eq.~\eqref{masterequation}.

\section{Single exponential potentials}
Since at large distances in field space string-theoretic scalar potentials tend to be exponential fall-offs, let us briefly recall the FLRW- and DW-type solutions for systems described by the following truncation (describing asymptotic behavior) in dimension $D>2$:
\begin{equation}
    L = T - V = - \tfrac{1}{2}(\partial\phi)^2 - \Lambda \, \e^{- \gamma\phi}.
\end{equation}
As pointed out in ref.~\cite{Skenderis:2006jq}, one can obtain static flat DW-flow solutions from FLRW solutions and vice versa by flipping the sign of $V$.
Hence, for we now restrict to FLRW solutions since the extension to DW flows is analogous, and we will come back to the correspondence in subsec. \ref{ssec.: FLRW/DW correspondence}.

Let us first consider positive potentials ($\Lambda>0$), and work in the gauge $N=1$. One can argue that, when
\begin{equation}
\gamma^2 < \gamma^2_c(D) \equiv 4 \, \frac{D-1}{D-2}\,,
\end{equation}
the late-time attractor is the \emph{scaling} solution \cite{Copeland:1997et}
\begin{subequations}
\begin{align}
\phi(\tau) & = \dfrac{2}{\gamma} \, \ln \, \biggl( \dfrac{\tau}{\tau_*} \biggr) - \dfrac{1}{\gamma} \, \ln \biggl[ (\gamma_{\mathrm{c}}^2 - \gamma^2) \dfrac{2}{\gamma^4} \dfrac{1}{\Lambda \tau_*^2} \biggr] \,, \label{scaling solution} \\
a(\tau) & = a(\tau_*) \biggl( \dfrac{\tau}{\tau_*} \biggr)^{\frac{4}{D-2} \frac{1}{\gamma^2}}\,,
\end{align}
\end{subequations}
where $\tau_* > 0$ is an arbitrary time. The name ``scaling'' arises because of two reasons.
The FLRW metric has a homothetic Killing vector, and, more importantly, the ratio of the kinetic energy over the potential energy is fixed: both terms scale in the same way with time.
In this case, we can check that the ``extended distance'' between $\phi(\tau_0) = \phi_0$ and $\phi(\tau)$ along the on-shell trajectory reads 
\begin{equation} \label{extended distance - cosmological solution for exponential potential}
\Delta = \dfrac{\gamma_{\mathrm{c}}}{\gamma} \, (\phi - \phi_0).    
\end{equation}
Noticeably, $\Delta$ is independent of the scale $\Lambda$ because the distance is invariant under constant field shifts, which just rescale $\Lambda$.
On the contrary, if $\gamma^2 \geq \gamma^2_c$, the late-time solution corresponds to \emph{kination} and it reads
\begin{subequations}
\begin{align}
\phi(\tau) & = \dfrac{2}{\gamma_{\mathrm{c}}} \, \ln \, \biggl( \dfrac{\tau}{\tau_*} \biggr) + \phi(\tau_*) \,, \label{kinating scaling solution} \\
a(\tau) & = a(\tau_*) \biggl( \dfrac{\tau}{\tau_*} \biggr)^{\frac{1}{D-1}}\,.
\end{align}
\end{subequations}
The name originates from the fact that the kinetic energy dominates over the potential energy at large times.
In this case, $\Delta = \phi - \phi_0$ and we recover exactly the ordinary distance conjecture, as expected.

Scaling solutions are a prototypical string-theoretic solution, since exponential potentials are ubiquitous \cite{Ooguri:2018wrx, Hebecker:2018vxz} and runaway solutions where the fields evolve asymptotically are generally encountered.
We observe that such scaling solutions indeed fulfill a fundamental property that we already discussed above: they evolve towards the asymptotics with no memory of the initial conditions.

When the potential instead is negative ($\Lambda<0$), it is sensible to look at contracting universes.
In this case, the scaling (kination) attractor arises in the opposite regime $\gamma^2 > \gamma^2_c$ ($\gamma^2 \leq \gamma^2_c$).
The explicit solutions are still given by the same expressions as above, with $\tau < \tau_* < 0$.
The solution describes a contracting universe that reaches a crunch as $\tau \to 0^-$.

Interestingly, the scaling solutions arise exactly in the regime for $\gamma$ where the pseudo-superpotential $\tilde{W}$ is an exponential.
In other words,
\begin{equation}
\tilde{W} = \tilde{\Lambda} \, \e^{- \frac{1}{2} \, \gamma \phi}
\end{equation}
for $\smash{\Lambda \gtrless 0}$ if $\smash{\gamma^2 \lessgtr \gamma^2_c}$, since $\Lambda = (\gamma^2_c-\gamma^2) \, \tilde{\Lambda}^2/8$. We therefore expect that the attractors in the opposite regimes are not pseudo-SUSY in the restricted sense of ref.~\cite{DiazDorronsoro:2016rrz}. Note that scaling solutions provide the on-shell behavior $\smash{\tilde{W} = 4/(\gamma^2 \tau)}$; for kinating solutions, we can just formally replace $\gamma$ with $\gamma_{\mathrm{c}}$.

\section{Testing the conjecture}
Let us discuss the generalized notion of field-space distance in simple examples.

\subsection{(A)dS space on a circle}
Consider an (A)dS$_{d+1}\times \mathrm{X}^{9-d}$ vacuum of string theory, so that we can truncate the $(d+1)$-dimensional description just to gravity and a cosmological constant:
\begin{equation}
S = \int \de^{d+1} x \, \sqrt{-g_{d+1}} \; \bigl( \tfrac{1}{2} \mathcal{R}_{d+1} - \Lambda \bigr)\,.
\end{equation}
We can reduce this theory over a circle of radius $l=1$:
\begin{equation}
ds^2_{d+1} = \e^{- \frac{2 \varphi}{\sqrt{d-1} \sqrt{d-2}}} \, ds^2_d + \e^{2 \varphi \frac{\sqrt{d-2}}{\sqrt{d-1}}} \, \de \theta^2\,.    
\end{equation}
Here $\theta$ is the circle angle and $\varphi$ is the canonical Einstein-frame radion.
The reduced theory then allows for the following consistent truncation:
\begin{equation}
S = \int \de^{d} x \, \sqrt{-g_{d}} \; \bigl[ \tfrac{1}{2} \mathcal{R}_{d} - \tfrac{1}{2}(\partial\varphi)^2 - \Lambda \, \e^{- \frac{2 \varphi}{\sqrt{d-1} \sqrt{d-2}}} \bigr]\,.
\end{equation}
So, we have a theory with an exponential potential of the type considered in the previous section, with an exponential coupling below the critical value, i.e. $\gamma^2 < \gamma_{\mathrm{c}}^2$.
Hence, we have a scaling solution for dS-reductions and kination for AdS-reductions for cosmology and the opposite for domain-wall solutions (see ssec. \ref{ssec.: FLRW/DW correspondence} for more details).
For instance, the domain-wall viewpoint for $\Lambda < 0$ in the gauge $N=1$ gives
\begin{equation}
\tilde{W} = \dfrac{(d-1)(d-2)}{\tau},
\end{equation}
for which the $\Lambda$-dependence drops out. This is a great consistency check because our distance measure is then invisible to $\Lambda$. To see it more explicitly, we can denote the starting and final scalar values to be $\varphi_{i,e}$. We find
\begin{equation}
\Delta = (d-1) (\varphi_e -\varphi_i) \,.
\end{equation}
We notice that our new definition of distance, in this example, is related to standard distance up to an order-1 (dimension-dependent) multiplicative factor.
How can we interpret this?

Consider for instance an $\mathrm{AdS}_5\times \mathrm{X}^5$ vacuum in string theory, say for example with $\mathrm{X}^5=\mathrm{S}^5$ in IIB string theory.
With the above procedure we would find a theory in 4 dimensions with negative exponential.
If we remove the $F_5$-flux and change $\mathrm{S}^5$ for $\mathbb{T}^5$ we find $\Lambda=0$.
Yet, the way the masses of the winding strings tower or KK tower (from the subsequent circle reduction) is identical for both cases and hence independent of $\Lambda$.
For the case without $\Lambda$, we can use the standard distance conjecture and the tower mass should be identical to the one where we include the potential.
We see that this is correct on the condition that we change the value of the exponent $\alpha$ in the distance formula $m(\Delta) = m(0) \, \e^{- \alpha \Delta}$ by an order-1 magnitude for scaling solutions (not for kinating ones).
As we said, this is harmless since the distance conjecture never postulates that $\alpha$ is a universal number: it changes from example to example, but has a conjectured a universal lower bound \cite{Etheredge:2022opl}.

\subsection{Flowing along the volume/dilaton}
Any geometric flux compactification features at least the volume scalar $\varphi$ of the extra dimensions and the string theory dilaton $\phi$.
If we look for asymptotic limits that correspond to weak coupling and/or large volume (potentially after using dualities), the scalar potential is dominated by ``classical effects" such as flux energies, brane tensions and Ricci curvatures of the extra dimensions \cite{VanRiet:2023pnx}.
In canonically-normalized variables, these energy contributions are all exponentials in these fields:
\begin{equation}
V = \sum_i \Lambda_i \, \e^{a_i \phi + b_i \varphi}\,.
\end{equation}
Here the dependence on other fields is assumed to be in the symbols $\Lambda_i$.
The numbers $a_i, b_i$ are all order-1 numerical factors.
As the volume grows large, the theory decompactifies and we expect the usual consistency with the ordinary distance conjecture with the KK modes becoming light, coherently with the fact that $V$ depletes in that limit.
Since attractor solutions are of either the scaling or the kination type, \emph{both} the ordinary distance conjecture and the generalized distance conjecture presented in this paper will be obeyed.
To see this, notice that for kination-type attractors $T+V$ reduces to $T$ and for scaling attractors $T + V = (1+1/\sigma) \, T$, with $\sigma > 0$ a finite constant.
In fact, this is a central conclusion of this work:
\begin{itemize}[leftmargin=*]
    \item[] \emph{In asymptotic limits that have a weak-coupling interpretation, the scalar potentials tend to be exponential and the corresponding scaling or kination attractor flows imply that a field-space distance formulated in the standard way, using only kinetic energy, is proportional to the generalized distance that includes the potential.}
\end{itemize}
One might worry that this conclusion is not general since we only constructed attractor solutions for a single potential, but is has been shown that attractor solutions for multiple exponentials obey the same structure in full generality, independently of initial conditions \cite{Shiu:2023fhb}.\footnote{This observation seems robust against the assumption of a flat field space. Evidence so far suggests the existence of linearly-stable cosmological solutions where the scalar kinetic energy is proportional to all other kinetic terms (including axions) and exponential potential terms, on general grounds. For instance, see refs.~\cite{Sonner:2006yn, Cicoli:2020cfj, Revello:2023hro}. Beyond linear stability, proofs for certain coupling configurations appear in ref.~\cite{Shiu:2024sbe}. Analogous considerations hold in the presence of negative potentials too \cite{Collinucci:2004iw, Hartong:2006rt}.}
In particular, even for multiple fields in a flat target space with multiple exponential potential, the attractor is 1-dimensional in field space: a linear combination of the scalars takes exactly the form in eqs. (\ref{scaling solution}, \ref{kinating scaling solution}), while all the other fields are constant.
All our previous considerations apply unchanged.

One might expect that the tower masses in asymptotic limits are not too different from the case without fluxes, since for example the mass operator for KK modes changes but the overall scaling laws are not affected.
Let us illustrate this with a flow along the $\mathrm{S}^5$-volume scalar in $\mathrm{AdS}_5 \times \mathrm{S}^5$. The compactification Ansatz is
\begin{equation}
ds^2_{10} = \e^{- \sqrt{\frac{5}{6}} \varphi} \, ds^2_5 + \e^{\sqrt{\frac{3}{10}} \varphi} \, \de \Omega_5^2\,,
\end{equation}
where $\de \Omega_5^2$ is the metric on $\mathrm{S}^5$.
The scalar potential has a piece from the curvature of the internal metric and one from the $F_5$-flux:
\begin{equation}
    V = - K_{\mathrm{S}^5} \, \e^{- \sqrt{\frac{32}{15}} \varphi} + N^2 \, \e^{-\sqrt{\frac{40}{3}} \varphi} \,,
\end{equation}
where $\smash{K_{\mathrm{S}^5}}$ is a positive constant proportional to the sphere curvature and $N$ is a number proportional to the 5-form flux.
We will not investigate the small-volume limit since we have no control over the scalar potential then. Instead we are interested in the large-volume regime, away from the $\mathrm{AdS}_5$-minimum. Then the curvature term depletes less quickly and we can approximate
\begin{equation}
    V \simeq - K_{\mathrm{S}^5} \, \e^{- \sqrt{\frac{32}{15}} \varphi} \,,
\end{equation}
which is an exponential potential with $\gamma^2=32/15<16/3=\gamma^2_c$.
The cosmological attractor is then a contracting kinating solution such that the generalized distance formula gives the same as the ordinary distance formula, which is obviously consistent with the ordinary KK-tower coming down in this limit.

\subsection{Have we generalized the distance conjecture?}
The original distance conjecture \cite{Ooguri:2006in}, formulated for pure moduli spaces, measures distances using the metric from the kinetic-energy term \emph{and} instructs one that the distance between two points has to be computed along the shortest geodesic connecting them.
A geodesic curve is also what the scalar fields track in an actual cosmological solution in the absence of a potential.
We have reviewed that weakly-coupled string theory, which is what typically arises in the asymptotic field regime (after choosing the right duality frame), has cosmological attractors that either are kinetic-energy-dominated (kination), or where the kinetic energy scales as the potential energy (scaling), even in the multi-field case.
In flat field spaces, this implies that our generalized distance is valid together with the original one that does not contain the scalar potential.
Does this mean that there is no additional information in the generalized distance conjecture?
We detail below that this is not the case.

A basic simple reason is that, often for a multi-field scaling solution, the asymptotic on-shell field-space trajectory is a straight line in a specific direction, determined uniquely by the potential and independently of the initial conditions \cite{Shiu:2023fhb}. This is \emph{one} geodesic trajectory in field space. So, our notion of distance is more restrictive than the distance conjecture with pure moduli in that we identify a precise geodesic.

Furthermore, there is one obvious point of potential greater difference: in case we have a potential other than a multi-exponential one and/or a curved field-space metric, which might well remain the case even in the asymptotics, there is no obvious reason why the cosmological solution should be a field-space geodesic.
For instance, one can take the prototypical axion/saxion model of two scalars $\phi$ and $\zeta$ with the Lagrangian density
\begin{equation} \label{saxion-axion model}
    L = - \dfrac{1}{2} \, (\der \phi)^2 - \dfrac{1}{2} \, \e^{- \lambda \phi} \, (\der \chi)^2 - \Lambda \, \e^{- \gamma \phi}.
\end{equation}
Depending on the relative size and sign of $\gamma$ and $\lambda$, there is evidence that this theory admits late-time attractors \cite{Sonner:2006yn, Sonner:2007cp, Cicoli:2020cfj} that are not geodesics \cite{Chemissany:2007fg, DiazDorronsoro:2016rrz}.

We can add more and relate this to our discussion of pseudo-/fake SUSY. In the literature, it has been claimed that all scaling solutions are geodesic \cite{Karthauser:2006ix}.
However, it was pointed out in ref.~\cite{Chemissany:2007fg} that this is not necessarily the case, with e.g. the scaling solution of refs.~\cite{Sonner:2006yn} providing an explicit counter example.
In ref.~\cite{DiazDorronsoro:2016rrz}, it was understood how this comes about and since the explanation matters for our purposes we briefly recall this here.
The main observation of ref.~\cite{Karthauser:2006ix} is that the scaling is argued to imply that scalars trace out curves along Killing vector fields (we use the gauge $N=1$):
\begin{equation}
    \nabla_{(i}\dot{\phi}_{j)} =0\,,
\end{equation}
where we lowered indices with $G_{ij}$. In case there is a pseudo-superpotential $\tilde{W}$, the tangent velocity 1-form is a gradient function
\begin{equation}\label{flow}
 \dot{\phi}^i =  -G^{ij} \partial_j \tilde{W}\,,
\end{equation}
such that the anti-symmetric derivative $\smash{\nabla_{[i}\dot{\phi}_{j]}}$ is zero as well, leading to the geodesic equation: $\smash{\dot{\phi}^j\nabla_{i}\dot{\phi}_{j}=0}$. 
If one assumes that the argument leading to the Killing flow holds water, it seems like the conclusion is inevitable since Hamilton-Jacobi theory guarantees the local existence of the $\tilde{W}$-function.
Yet there is a subtlety.
The flow equation \eqref{flow} does not guarantee that the velocity can be thought of as a vector field, which is a necessary condition for the derivation to make sense.
What Hamilton-Jacobi theory tells us is that $\tilde{W}$ is a function of $n$ integration constants, where $n$ is the number of scalars.
However, these constants can be initial velocities, positions or mixtures of the two.
It is only when the integration constants are initial velocities that can one interpret the 1-form $\de \tilde{W}$ as an element of the co-tangent bundle.
This special condition was named ``restricted pseudo-/fake SUSY'' \cite{DiazDorronsoro:2016rrz}.
A famous example of a flow that fails to be pseudo-SUSY in this restricted sense is that of a free scalar, where the function $\tilde{W}$ depends on the initial position \cite{DiazDorronsoro:2016rrz}. The same applies to the scaling solution found in ref.~\cite{Sonner:2006yn} and it explains why that solution is not a geodesic.\footnote{Since (a difference in) superpotential functions $\tilde{W}$ relates to tensions of domain walls, the link between first-order flows for scaling solutions and geodesics explained above could very well be related to the observations in ref.~\cite{Etheredge:2023zjk}.} 

To sum up, our generalized distance conjecture, in asymptotic regimes, does not necessarily involve geodesic trajectories. Hence, more outspoken differences with the ordinary distance measure can occur.

Regarding the trajectory of integration, another difference with the original distance conjecture is about the ability of patching the field space. Contrary to the case of pure moduli in ref.~\cite{Ooguri:2006in}, our proposal is only defined to compute distances between points that belong to an attractor trajectory, which is generally a lower-dimensional subspace of the full field space. As an example, the field space for canonical scalars is the Euclidean space, but in the presence of exponential potentials the attractor is only a straight line within it.

\subsection{Is there an ambiguity?} \label{ssec.: FLRW/DW correspondence}
Our prescription for a generalized distance employs two related but distinct possible classes of solutions: FLRW- and DW-type solutions ($\epsilon=-1$ and $\epsilon=+1$, respectively).
Hence, one should rightfully wonder whether there is a physically-motivated distinction.

This puzzle has already been mentioned in the study of canonical scalars with exponential potentials with slope $\gamma$ and we saw that the exponential dependence of a mass gap on the distance persists, whether it be computed for $\epsilon=-1$ or $\epsilon=+1$.
However, one might fear that the generalized notion of distance is multi-valued.
Although it is true that for a given $\gamma$  we compute two different distances, $\smash{\Delta_{\mathrm{FLRW}}}$ and $\smash{\Delta_{\mathrm{DW}}}$ respectively, these only differ up to a constant factor $\gamma_{\mathrm{c}}/\gamma$.
For pure moduli spaces (without a scalar potential), we expect the tower of states to become light exponentially as in eq. (\ref{distance}), where the constant $\alpha$ is not fixed even for the well-established notion of distance in pure moduli spaces.\footnote{There are conjectured and well-tested expectations \cite{Etheredge:2022opl}, but the key property is that the tower of states becomes light exponentially in the distance.}
In the case of field spaces with a scalar potential, therefore, we simply expect that a relationship of the form 
\begin{equation} \label{FLRW/DW equivalence}
    \smash{\alpha_{\mathrm{FLRW}} \Delta_{\mathrm{FLRW}} = \alpha_{\mathrm{DW}} \Delta_{\mathrm{DW}}}   
\end{equation}
will hold, for two different order-1 constants $\smash{\alpha_{\mathrm{FLRW}}}$ and $\smash{\alpha_{\mathrm{DW}}}$.
In fact, one can even do more.
The $\gamma$-parameter choices of kinating ($\mathrm{K})$ and scaling ($\mathrm{S}$) types are complementary to each other in the two signatures \cite{Sonner:2005sj, Skenderis:2006jq}:\footnote{This statement is true as long as one fixes the conserved quantity $\delta = \mathrm{sgn} \, (\dot{a}/a)$ as $\delta = -\epsilon$. Throughout this paper, we always fix $\delta = -\epsilon$. Without this choice, additional bifurcations emerge in patching the parameter space, but conceptually the whole discussion holds intact.} for any value of $\gamma$, there exists a signature $\epsilon$ such that the asymptotic attractor is of the kinating type, as displayed in fig. \ref{fig.: saxion + potential}.
Hence, one might simply define the generalized distance with the prescription of picking the $\epsilon$-sign which implies a kinating solution.
It is then crucial to observe that kinating solutions can span any field-space point at infinite distance because they are insensitive to the potential.
In this case, the ordinary and the generalized distances match identically and, on top of that, the attractor curves are geodesics.

\begin{figure}[H]
\centering
    
\begin{tikzpicture}[scale=0.95,dot/.style={draw,circle,minimum size=1.65mm,inner sep=0pt,outer sep=0pt,black,fill=magenta,solid}]

    \draw[-, ultra thick, orange] (0,0) node[below, black] {$0$} -- (3.25,0) node[above,black,pos=0.5,align=center]{$\epsilon=-1$: $\mathrm{S}$ \\[2pt] $\epsilon=+1$: $\mathrm{K}$ \\[-4pt]};
    \draw[->, ultra thick, teal] (3.25,0) -- (6.5,0) node[below, black]{$\gamma$} node[above,black,pos=0.5,align=center]{$\epsilon=-1$: $\mathrm{K}$ \\[2pt] $\epsilon=+1$: $\mathrm{S}$ \\[-4pt]};
    
    \node[dot] at (3.25,0){};
    \node[below, align=center] at (3.25,0){\\[-3pt] $\gamma_{\mathrm{c}}$};

    \end{tikzpicture}
    
\caption{Asymptotic attractors for canonical scalars in an exponential potential with slope $\gamma$.}
    
\label{fig.: saxion + potential}
\end{figure}

This structure extends well beyond a single scalar with a single exponential potential.
For instance, multiple scalar models with flat target spaces and multi-exponential potentials generalize like this as well \cite{Shiu:2023nph, Shiu:2023yzt}.
Hence, for this broad class of models, we can argue for a non-ambiguous distance proposal which is identical to the conventional distance conjecture for moduli spaces.

Noticeably, there are counter-examples for curved target spaces where the attractors are non-geodesic.
For instance, in the axion/saxion models that we already introduced in eq. (\ref{saxion-axion model}), it is not generally true that at any point in parameter space the asymptotic solution is kination for at least one sign $\epsilon$.
As a consequence, not every point at ``infinite distance'' can be probed.
Yet, all linearly-stable asymptotic solutions are such that the kinetic energy scales like the scalar potential, i.e. they fulfill eq. (\ref{FLRW/DW equivalence}).
To illustrate these points, we refer to the explicit saxion/axion model.
For any given point in the parameter space $(\lambda, \gamma) \in \mathbb{H} = \mathbb{R}^+ \times \mathbb{R}$, fixing e.g. $\epsilon=-1$, one can individuate the linearly-stable asymptotic solutions.
The kinating solution is the linearly-stable solution in the region $\smash{\mathbb{K}_{-1} = \lbrace (\lambda, \gamma) \in \mathbb{H}: \lambda \leq 0 \wedge \gamma \geq \gamma_{\mathrm{c}} \rbrace}$, while the linearly-stable solutions in the complementary space $\smash{\mathbb{H} \smallsetminus \mathbb{K}_{-1}}$ are either of the scaling type or the non-geodesic ($\mathrm{NG}$) solutions mentioned before.
For the opposite signature $\epsilon=+1$, however, the linearly-stable solution is not kination throughout such a complementary space, as displayed in figs. \ref{fig.: axion/saxion + potential, FLRW}\footnote{A similar but different plot appears in ref. \cite{Cicoli:2020noz}. The differences originate in the fact that in the current paper we are not considering an additional barotropic fluid besides the scalars.} and \ref{fig.: axion/saxion + potential, DW}.

\begin{figure}[H]
\centering
    
\begin{tikzpicture}[scale=0.95,dot/.style={draw,circle,minimum size=1.65mm,inner sep=0pt,outer sep=0pt,black,fill=magenta,solid}]

    \hfill;

    \fill[orange!15!white] (-3.2,0) rectangle (3.2,1.5);
    \fill[teal!15!white] (-3.2,1.5) rectangle (0,3.0);
    
    \begin{scope}
        \fill[cyan!10!white] (0,1.5) rectangle (3.2,3.0);
        \clip (4,1.5) ellipse (4 and 1.2);
        \fill[cyan!10!white] (0,0) rectangle (3.2,3.2);
    \end{scope}

    \draw[->] (-3.5,0) -- (3.5,0) node[below]{$\lambda$};
    \draw[->] (0,-0.5) -- (0,3.5) node[left]{$\gamma$};
    
    \begin{scope}
        \clip (-1,0) rectangle (3.2,1.5);
        \draw[thick, magenta] (4,1.5) ellipse (4 and 1.2);
    \end{scope}
    
    \draw[thick, magenta] (-3.2,1.5) -- (0,1.5);
    \draw[thick, magenta] (0,1.5) -- (0,3.0);

    \node[] at (-1.6,2.25){$\mathbb{K}_{-1}$};
    \node[] at (1.6,1.5){$\mathbb{NG}_{-1}$};
    \node[] at (-1.6,0.75){$\mathbb{S}_{-1}$};
    \node[below left] at (0,0){$0$};
    \node[below left] at (0,1.5){$\gamma_{\mathrm{c}}$};

    \end{tikzpicture}
    
\caption{Asymptotic linearly-stable solutions for canonical scalars in an exponential potential with slope $\gamma$ and a kinetic axion coupling $\lambda$, when $\epsilon=-1$. The curve separating scaling and non-geodesic solutions is $\lambda = \gamma_{\mathrm{c}}^2/\gamma - \gamma > 0$.}
    
\label{fig.: axion/saxion + potential, FLRW}
\end{figure}

\begin{figure}[H]
\centering
    
\begin{tikzpicture}[scale=0.95,dot/.style={draw,circle,minimum size=1.65mm,inner sep=0pt,outer sep=0pt,black,fill=magenta,solid}]

    \fill[teal!15!white] (0,0) rectangle (3.2,1.5);
    \fill[teal!15!white] (-3.2,0) rectangle (0,3.0);
    \fill[orange!15!white] (0,1.5) rectangle (3.2,3.0);

    \draw[->] (-3.5,0) -- (3.5,0) node[below]{$\lambda$};
    \draw[->] (0,-0.5) -- (0,3.5) node[left]{$\gamma$};
    
    \draw[thick, magenta] (0,1.5) -- (3.2,1.5);
    \draw[thick, magenta] (0,1.5) -- (0,3.0);

    \node[] at (1.6,2.25){$\mathbb{S}_1$};
    \node[] at (-1.6,0.75){$\mathbb{K}_1$};
    \node[below left] at (0,0){$0$};
    \node[below left] at (0,1.5){$\gamma_{\mathrm{c}}$};

    \end{tikzpicture}
    
\caption{Asymptotic linearly-stable solutions for canonical scalars in an exponential potential with slope $\gamma$ and a kinetic axion coupling $\lambda$, when $\epsilon=1$.}
    
\label{fig.: axion/saxion + potential, DW}
\end{figure}

This single example proves the possible absence of asymptotic kination for certain patches in parameter space.
This point deserves further investigation.
We now discuss a similar model -- obtained from a top-down construction -- that also features non-geodesic attractors and that seems to contradict our generalized notion of distance.

\subsection{Are there counter-examples?}
A KK-tower that does not depend exponentially on the field-space distance -- computed along the on-shell trajectory -- was observed in ref.~\cite{Buratti:2018xjt} for warped throats of the Klebanov-Strassler type \cite{Klebanov:2000hb}.\footnote{We thank José Calderón-Infante for bringing this article to our attention.}
In detail, ref. \cite{Buratti:2018xjt} identifies a 5d EFT for an axion $\theta$ and the dilaton $\phi$ such that, on the on-shell trajectory $\mathrm{C}$, the KK tower $\smash{m_{\mathrm{KK}} = m_{\mathrm{KK}, 0} \, \e^{-16 \omega/3}}$, where $\omega$ is the Einstein-frame radion, falls off polynomially in the moduli-space distance
\begin{equation}
    \tilde{\Delta} = \int_{\mathrm{C}} \de \tau \, \sqrt{G_{ij} \dot{\phi}^i \dot{\phi}^j},
\end{equation}
rather than exponentially.\footnote{One should note that the EFT in ref. \cite{Buratti:2018xjt} is defined as the compactification of a consistent truncation including the volume $\omega$.
Such an EFT is defined for scales far below the mass at which $\omega$ is stabilized, which is also the mass scale of the truncated-out modes, but it includes the backreaction on $\omega$ itself due to the axion dynamics, replacing $\omega$ by $\omega(\theta, \phi)$ in the action.}
As the trajectory does not correspond to a geodesic in the $(\theta, \phi)$-space, this was argued to be in no contradiction with the original distance conjecture.
One might hope that the KK tower turned out to depend exponentially on the generalized distance $\Delta$, which would represent a clear example going beyond the original moduli-space distance, but unfortunately this is not the case.
One can see that the KK mass scales as
\begin{equation}
    m_{\mathrm{KK}} = \dfrac{\mu}{\Delta^{2/3}} = \dfrac{\nu}{\tilde{\Delta}^{4/3}},
\end{equation}
for two constants $\mu$ and $\mu$.
Hence, this model poses a question mark on the generality of our proposed generalized distance and deserved further investigation.

\subsection{Dimensional reduction}
An important property of the Swampland conjectures is preservation under dimensional reduction; see e.g. refs. \cite{Heidenreich:2015nta, Heidenreich:2019zkl, Rudelius:2021oaz, Etheredge:2022opl, vandeHeisteeg:2023dlw}. One typically finds $D$-dependent constants appearing in bounds and the dependence on $D$ is conserved in any dimension $D>2$. We can confirm the same structure for our notion of extended distance.

According to the strong dS conjecture \cite{Rudelius:2021oaz, Rudelius:2021azq}, any potential for a single field $\phi$ should have a coupling $\gamma(D) \geq 2/\sqrt{D-2}$.\footnote{See also the arguments in refs.~\cite{Shiu:2023fhb, Hebecker:2023qke} for discussions beyond the single-field case.} Hence, the distance of eq. (\ref{masterequation}) for a $D$-dimensional cosmological solution with an exponential potential should be bounded as
\begin{equation} \label{dimensional reduction}
    \Delta \leq \sqrt{D-1} \, (\phi - \phi_0).
\end{equation}
Upon reduction down to dimension $d=D-k$, a canonical radion $\varphi$ is introduced and the $d$-dimensional potential is
\begin{equation}
    V_d = \Lambda \, \e^{- \frac{2 \sqrt{k} \, \varphi}{\sqrt{d+k-2} \sqrt{d-2}} - \gamma(D) \phi}.
\end{equation}
By a basis transformation, the potential is actually an exponential for a single field, with a coupling $\gamma(d) \geq 2/\sqrt{d-2}$.
A second field is a pure modulus and in the asymptotics it is always frozen.
If the compact space has negative curvature, the multi-field potential along the cosmological solution is equivalent to a single potential for a single field $\sigma$, still with a coupling $\gamma(d) \geq 2/\sqrt{d-2}$ \cite{Collinucci:2004iw, Shiu:2023nph}.
In all cases, the dimension dependence of the bound on $\Delta$ is preserved as in eq.~(\ref{dimensional reduction}).
For negative potentials, one automatically has $\Delta \leq (\phi - \phi_0)$ and on-shell solutions for reductions with multiple fields can be constrained, too, demonstrating the same result \cite{Shiu:2023yzt}. Consequently, for all DW-solutions, the inequality is in place as well.

\subsection{Generalized distance and Planck units}
A natural question that one might ask is how the distance changes in theories where the flow along a field-space trajectory changes the Planck scale.
This is a very common scenario in string compactifications, where for instance the dilaton and the volume change across different vacua or along a dynamical solution.

The Einstein-frame metric is defined up to an arbitrary constant, but changing such constant does not change the relative ratios of scales in the new units one must use.
For instance, if we write the gravitational coupling $\kappa_d$ explicitly, the product $\kappa_d \phi$ that appears in the ordinary moduli-space distance is independent of any constant rescaling.
Similarly, the addition of a potential $V$ does not induce any change, since in any frame one must just express the potential in the appropriate units.
Hence, our generalized distance is robust against changes of units.

\section{An application to ekpyrosis}
In UV-complete effective field theories coupled to gravity, positive potentials with small slopes in asymptotic regions of field space are in tension with the (refined) dS conjecture \cite{Obied:2018sgi, Ooguri:2018wrx}.
One set of arguments for this relies on the distance conjecture in configuration space \cite{Ooguri:2006in}.
Similarly, one might wonder whether Swampland principles suggest that quantum gravity also abhors the exact opposite, i.e. negative and very steep potentials \cite{Bernardo:2021wnv, Andriot:2022brg}.
In fact, such a kind of potentials does not seem immediate to engineer for string compactifications in the regime of classical supergravity \cite{Uzawa:2018sal, Shiu:2023yzt}.
Besides the formal interest on whether such potentials exist in the string landscape, they are also of phenomenological interest for the so-called \emph{ekpyrotic scenario} of the early universe \cite{Khoury:2001wf, Khoury:2001bz, Steinhardt:2002ih}.
This is an alternative proposal to the theory of cosmic inflation which could solve the same fine-tuning problems related to the early universe by relying on a phase of cosmic contraction instead of very rapid expansion.
Scalar fields in steep and negative potentials provide an effective theory for such a contraction.

Based on the distance conjecture, ref.~\cite{Hebecker:2018vxz} provided a particularly simple argument as to why positive potentials in string compactifications should fall off exponentially near the moduli-space boundary.
In essence, the fact that a tower of states gets exponentially light lowers the cutoff $\Lambda_{\mathrm{s}}$ of the theory accordingly. For an effective theory to make sense, we must require $H<\Lambda_{\mathrm{s}}$.
As $H$ decreases, it is possible to remain within an effective theory throughout. Since $H^2$ is proportional to the total energy $T+V$, the argument follows through.
For a contracting universe, this reasoning is dramatically different since $H^2$ increases and one is forced out of the EFT quite quickly. Let us quantify all this.

By the distance conjecture, as one approaches the field-space boundary, a tower of states  of mass $m_n(\phi) = n^{\sigma} m(\phi)$ becomes light as
\begin{equation}
    m(\phi) \simeq m(\phi_0) \, \e^{- \alpha (\phi - \phi_0)}
\end{equation}
for some order-1 constant $\alpha$, with $n \in \mathbb{N}$ and $\sigma>0$ a constant depending on the microscopic origin of the tower (e.g. $\sigma=1$ for KK-states and $\sigma=1/2$ for string excitations). On the other hand, we can identify the cutoff with the species scale $\smash{\Lambda_{\mathrm{s}} = N^{-1/(d-2)}}$. If the degeneracy of each mass level is approximately constant,\footnote{This is always a valid approximation if the scale $m(\phi_0)$ is sufficiently close to the Planck mass. We thank Timm Wrase for discussing this point.} then the species scale behaves as
\begin{equation} \label{species scale}
    \Lambda_{\mathrm{s}} \simeq B \, \e^{- \frac{\alpha (\phi - \phi_0)}{(d-2) \sigma + 1}},
\end{equation}
for a constant $B>0$. Evidence for this behavior has recently been confirmed in various ways \cite{vandeHeisteeg:2022btw, vandeHeisteeg:2023dlw, vandeHeisteeg:2023ubh, Cribiori:2023sch}. For an expanding FLRW-universe with a positive potential $V>0$, because $H^2 \geq 2 V/[(d-1)(d-2)]$, the requirement that $\Lambda_s > H$ bounds the behavior of the scalar potential to be of the form $\smash{V(\phi) < A_+ \, \e^{- 2\alpha (\phi - \phi_0)/[(d-2) \sigma + 1]}}$, for a constant $A_+>0$. A similar argument follows if we replace the field displacement by the general geodesic distance. The proportionality relation $T = \sigma V$ that we expect for on-shell asymptotic solutions supports these arguments also in view of the extended notion of distance, where at most only a change in the constant $\alpha$ should be expected.\footnote{We emphasize that the bound on $V$ is not only in place for slow-roll evolution, in which $T \ll V$.}

A contracting FLRW-universe behaves differently. Let $V \leq 0$.
By the Friedmann equations, the kinetic energy is bounded as $\smash{H^2 \leq 2 T/[(d-1)(d-2)]}$; moreover, we know that $\smash{\dot{H} = - 2 T/(d-2)}$. If $H(\tau_0)<0$, we can thus infer the inequality $\smash{H(\tau) \leq 1/[(d-1) \tau]}$, where $\tau<0$ and we fixed the gauge $\tau_0 = 1/[(d-1) H_0]$. Hence, we find the bound
\begin{equation} \label{ekpyrotic H-behavior}
    H^2(\tau) \geq \dfrac{1}{(d-1)^2 \tau^2}.
\end{equation}
In particular, this means that the Hubble scale diverges as $\tau \to 0^-$, and so does the kinetic energy $\smash{T \geq 2/(\gamma_{\mathrm{c}}^2 \tau^2)}$. For a single-field theory, in which $\smash{T = \dot{\phi}^2/2}$, we get the inequality
\begin{equation} \label{ekpyrotic phi-behavior}
    \phi(\tau) - \phi(\tau_0) \geq \dfrac{2}{\gamma_{\mathrm{c}}} \, \ln \, \dfrac{\tau_0}{\tau},
\end{equation}
in the convention where the field grows over time. The field displacement is always bounded from below by that of the purely-kinating universe. If we plug this into eq. (\ref{species scale}), we find that the species scale behaves as
\begin{equation}
    \Lambda^2_{\mathrm{s}}(\phi(\tau)) \lesssim B^2 \, \biggl( \dfrac{\tau}{\tau_0} \biggr)^{\!\! \frac{2}{(d-2)\sigma+1} \frac{\alpha}{\gamma_{\mathrm{c}}}}.
\end{equation}
Therefore, comparing with eq. (\ref{ekpyrotic H-behavior}), we see quantitatively the obstruction to keeping $\smash{\Lambda^2_{\mathrm{s}} \geq H^2}$ throughout the ekpyrotic phase. In particular, the Hubble parameter reaches the cutoff scale well before it would reach the Planck scale. In the simplest ekpyrotic universe, we need $\smash{-1 + 2T/(T+V) \geq w}$, where $w$ is a critical value $w \gg 1$. This means that we have $\smash{V \leq -[(w-1)/(w+1)] \, T}$, so we also find the constraint
\begin{equation}
    V(\phi(\tau)) < - A_- \, \e^{\gamma_{\mathrm{c}} (\phi(\tau) - \phi_0)},
\end{equation}
where $A_- > 0$ is a constant. Finding bounds on ekpyrosis using entropy arguments like ref.~\cite{Ooguri:2018wrx} did for positive potentials is also possible \cite{Bernardo:2021wnv}, but more involved. We leave a study of this in view of the extended distance conjecture for future work.

Thus far we discussed ekpyrosis towards the field-space boundary, but one can wonder what happens if moving towards the field-space bulk. In the appendix we point out that the Kallosh-Kachru-Linde-Trivedi (KKLT) potential \cite{Kachru:2003aw} prior to uplift might display all the necessary tools.
Even if the fields move from the boundary towards the bulk, in certain cases like that of ekpyrosis our generalized distance is still possible to compute. This is because there are attractor solutions bounding the evolution of the field-space trajectory.

\section{Discussion}
The distance conjecture has been central to much of the progress in the Swampland Program. However, to make it useful in a phenomenological context, it should be applied outside of the context of pure moduli spaces.
Hence, we are led to propose an extension of the conjecture that might be applicable to scalar fields that reside in a scalar potential.
Taking inspiration from how a potential in classical mechanics can be seen as due to integrating out free fields on a curved configuration space, and the extension thereof that includes gravity in a cosmological context \cite{Townsend:2004zp}, we came with a specific formula for the distance $\Delta$, in eq.~\eqref{masterequation}.
We suggest that this should be the distance controlling the falloff of the mass gap of the tower becoming exponentially light in the field-space asymptotics.

Any extended version of the distance conjecture can be tested against examples where we know the tower of massive particles along some field trajectory.
We believe the simplest possible example is obtained by reducing AdS space on a circle and our proposal holds in that case.
We then argued that it is expected to always hold in asymptotic regimes where string theory reduces to a classical compactification of 10d supergravity and, even more, that the ordinary distance conjecture is true \emph{at the same time}.
The reasons for that rest on the following three observations.
\begin{enumerate}
    \item In the asymptotic regime, the scalar potential has exponential fall-offs.
    \item Dynamical system theory applied to exponential potentials has shown that cosmological attractors in this case are either of the kination type or of the scaling type.
    \item When the attractor is of the kination type, our generalized distance formula becomes equal to the ordinary one. When the attractor is scaling, the generalized distance $\Delta$ is proportional to the original one with an overall order-1 coefficient. Since the coefficient $\alpha$ in the exponential that determines the tower mass is not a universal number, it means that both distance conjectures then apply.
\end{enumerate}
We furthermore observe two interesting relations with Hamilton-Jacobi theory and what has been named fake or pseudo-SUSY.
Firstly, the generalized distance equation can be rewritten as the integral over the (fake or pseudo-) superpotential.
Secondly, when the cosmological flow is fake-SUSY in the restricted sense \cite{DiazDorronsoro:2016rrz}, the cosmological attractors are geodesics \cite{DiazDorronsoro:2016rrz, Chemissany:2007fg} and the relation with the ordinary distance conjecture becomes very close since one would not only measure ordinary distance along a curve, but the flow also corresponds to a geodesic such that one ends up computing a geodesic distance. Yet, not all cosmological attractors are geodesics. 

This close connection between the ordinary distance conjecture and the generalized one implies that previous Swampland results based on the ordinary distance conjecture, yet applied to scalars with potentials, such as in ref.~\cite{Ooguri:2018wrx},
are qualitatively unchanged.

We formulated our proposal for a generalized distance in theories that admit attractor solutions to the field equations, evolving towards (or from) the field-space asymptotics.
In this sense, such a definition does not apply to computing distances between, for instance, two nearby vacua.
One can measure distances between AdS vacua by using a measure on metric space, through which the far-reaching AdS distance conjecture can be obtained \cite{Lust:2019zwm}.\footnote{For further interesting alternative suggestions to compute distances between AdS vacua, see refs.~\cite{Li:2023gtt, Palti:2024voy}.}
It has been shown that distances between AdS flux vacua can also be obtained by measuring distances in scalar field space, for fields that allow one to interpolate between these vacua. These scalar fields reside in a scalar potential with the various AdS minima \cite{Shiu:2022oti, Shiu:2023bay} (see also refs.~\cite{Farakos:2023nms, Tringas:2023vzn}).
However, the distances computed in refs.~\cite{Shiu:2022oti, Shiu:2023bay,Farakos:2023nms, Tringas:2023vzn} ignore the contribution from the potential to the distance.
Yet, we argued that the generalized distance conjecture is often equivalent to the ordinary one, up to overall coefficients, in the asymptotic regime.
It is not obvious to us that this also applies to the scalar fields that interpolate between AdS minima of refs.~\cite{Shiu:2022oti, Shiu:2023bay}.
In this case, one could try and employ our
generalized distance for flowing from one flux vacuum to a faraway different flux vacuum, effectively coarse-graining (averaging) over
the wiggles in the potential.
Potentially an outspoken difference with the ordinary distance conjecture is possible. We leave this for future research.

Finally, we applied our generalized distance conjecture to a situation which is seldom discussed in the Swampland program, i.e. steep and negative potentials.
Whereas we have not found a Swampland explanation by which such potentials cannot be found in asymptotic regimes of field space, our extended notion of distance implies at least that ekpyrotic phases in cosmology will leave the EFT regime more quickly than one naively expects, well before the Hubble density becomes Planckian.
To reach this conclusion, the calculation of the field-space distance in the presence of a potential is very much needed because the potential grows larger and larger.
We furthermore remark the straightforward yet previously-unobserved fact that the KKLT scalar potential, without the uplift term, might provide a stringy example of an ekpyrotic potential away from the asymptotic field regime.
\\

{\bf Note added.} While our work was in the final stages of completion, ref.~\cite{Mohseni:2024njl} appeared, with significant overlap with our results. We acknowledge that the inspiration to work on this problem came from discussions with the authors of this reference. We comment on the comparison between the extended notions of distance presented in this note and in ref.~\cite{Mohseni:2024njl} in app. \ref{app.: comparison}.

\begin{acknowledgments}
\subsection*{Acknowledgments}
We happily acknowledge discussions with Jake McNamara, Miguel Montero, John Stout and an anonymous referee. T.V.R. would like to thank the Harvard Swampland initiative for hospitality and the Leuven student encampment (KULStudents4Palestine) for providing an inspirational environment for research and more.

The research of T.V.R. is in part supported by the Odysseus grant GCD-D5133-G0H9318N of
FWO-Vlaanderen.
F.T. is supported by the FWO Odysseus grant GCD-D0989-G0F9516N.
\end{acknowledgments}

\appendix

\section{Ekpyrosis towards the field-space bulk}
If we do not include the anti-D3-brane uplift, the KKLT scenario \cite{Kachru:2003aw}
features a scalar potential for the universal Einstein-frame radion $\varphi$ of the form $\smash{V = V_+ + V_-}$, where we single out the positive- and negative-definite terms in an obvious notation, with
\begin{align*}
    V_+(\varphi) & = \Lambda \, \e^{- \sqrt{\frac{2}{3}} \varphi} \, \e^{- 2 a \, \e^{\sqrt{\frac{2}{3}} \varphi}}, \\
    V_-(\varphi) & = - K \, \e^{- \sqrt{\frac{8}{3}} \varphi} \, \e^{- a \, \e^{\sqrt{\frac{2}{3}} \varphi}}.
\end{align*}
All other fields are stabilized at higher energy scales and appear with their expectation values inside the Gukov-Vafa-Witten term $\ab W_0 \ab$, while $a$ and $\ab A \ab$ are constants depending on the details of the process of gaugino condensation responsible for breaking the no-scale structure of the theory. Here, the constants are $\Lambda = a^2 \ab A \ab^2/6$ and $K = a \ab A W_0 \ab/2$.
Leaving the physics of the dS uplift aside, remarkable evidence has been gathered for the existence of an AdS vacuum for the potential $V$ in flux compactifications with huge hierarchies with respect to the Planck scale \cite{Demirtas:2019sip, Demirtas:2021nlu, Demirtas:2021ote}.
Starting instead at very large volume, far from the AdS minimum, we can approximate the total potential as the steep falloff $V \simeq V_-$: the potential can evolve towards smaller volumes while being negative and extremely steep. Hence, we can model an ekpyrotic phase in terms of $V_-$.

We can estimate how well this potential might describe an ekpyrotic early universe. To tackle the problem, let us consider an effective theory with a negative-definite potential $V<0$. If there exist two constants $\Gamma_1 > \Gamma_2 > 0$ such that the gradient is bounded to have a bounded non-negative convexity as
\begin{equation}
    - \Gamma_2 V(\phi) \leq \dfrac{\der V}{\der \phi}(\phi) \leq - \Gamma_1 V(\phi),
\end{equation}
then we can show that the $w$-parameter of the theory falls in the constant window
\begin{equation}
    -1 + \dfrac{\Gamma_2^2}{3} \leq w \leq -1 + \dfrac{\Gamma_1^2}{3}.
\end{equation}
These hold at any time $\tau$ in the interval $\tilde{\tau} < \tau < \tau_\star$, where $\tilde{\tau}$ is some time along the dynamical evolution and $\tau_\star$ is the time of the big crunch. The idea is simply that the fixed convexity of the potential allows us to bound the time evolution in terms of two limiting exponential potentials, which are easier to study. As an application, let us consider a potential
\begin{equation}
    V(\phi) = - K \, \e^{-\beta \, \e^{\alpha \phi}},
\end{equation}
for $K, \alpha, \beta > 0$. So, if the field $\phi$ evolves over time from some value $\smash{\phi_1}$ to some value $\phi_2 < \phi_1$, we can identify $\smash{\Gamma_2 = \alpha \beta \, \e^{\alpha \phi_2}}$ and $\smash{\Gamma_1 = \alpha \beta \, \e^{\alpha \phi_1}}$.

For the potential $V_-$, neglecting the additional single-exponential dependence of the potential on $\smash{\varphi}$, one finds $\alpha = \sqrt{2/3}$ and $\beta = a$. Assuming the minimum radion value $\smash{\varphi_2}$ to be larger than the would-be KKLT vacuum minimum $\smash{\langle \varphi \rangle}$, if we can require the would-be volume to be $\smash{\langle \mathrm{vol}_6 \rangle = q \, l_s^6}$, with e.g. $q = 10^2$, then we need $\smash{\varphi_2 > \langle \varphi \rangle = \sqrt{8/3} \, \ln \, q}$. So, we get $\smash{\Gamma_2^2 > \sqrt{2 a^2/3} \, q^{2/3}}$.
This leaves pretty good flexibility on the choice of initial conditions $\varphi_1$ determining the bounding value $\Gamma_1$.
We should make sure that the ekpyrotic phase lasts for long enough as to solve the horizon problem, but also not too long as to make our assumption that we can trust just one term of the KKLT-potential, being far away from the minimum.
In order for an ekpyrotic phase to solve the horizon problem, we need $(a_2 H_2)/(a_1 H_1) \gtrsim 10^{28}$, where $\tau_1$ and $\tau_2$ are the times when ekpyrosis begins and ends, respectively. Based on the fixed convexity of the potential, we can see that
\begin{equation}
    \biggl( \dfrac{a_2 H_2}{a_1 H_1} \biggr)^{\!\! 2} \geq \biggl[ 1 - \dfrac{\Gamma_2^2}{\Gamma_1^2} \biggr]^{\frac{4}{\Gamma_2^2} - 2}.
\end{equation}
We can also check the extended distance conjecture. For a time evolution between a time $\tau_1$ and a time $\tau_2<0$, the same inequalities give
\begin{equation} \label{KKLT - extended distance}
    \dfrac{2 \sqrt{6}}{\Gamma_1^2} \, \ln \, \dfrac{\tau_1}{\tau_2} \leq \Delta \leq \dfrac{2 \sqrt{6}}{\Gamma_2^2} \, \ln \, \dfrac{\tau_1}{\tau_2}.
\end{equation}
This is bounded by the same logarithmic behavior in time as usual, but with prefactors that are inverse exponentials in canonical variables. Yet, the distance diverges at the big-crunch time $\tau_2 = 0^-$.

\section{Different generalized distances} \label{app.: comparison}
As we mentioned, a different notion of a field-space distance in the presence of a scalar potential was recently introduced in ref.~\cite{Mohseni:2024njl}.
Let us call this distance $\smash{\Delta_\MMVV}$.
Given two points $\phi_i$ and $\phi_e$ in field space, $\smash{\Delta_\MMVV}$ is defined as the tension of the domain wall connecting the two points, normalized by the total Euclidean energy density.
We can write
\begin{equation}
    \Delta_\MMVV = \int_{\phi_i}^{\phi_e} \de \phi \, \sqrt{1 - \dfrac{V_\E}{V_\E-T_\E}},
\end{equation}
where the subscript indicates the Euclidean configuration.
In multi-field settings, one must follow an Euclidean field configuration that interpolates the two endpoints while solving the field equations.
Rotating back to Lorentzian time $\tau_\E \rightarrow \I \tau$, we may write
\begin{equation}
    \Delta_\MMVV = \int \de \tau \, \sqrt{2 T} \sqrt{\dfrac{T}{T+V}},
\end{equation}
which is different from the proposed distance in eq.~(\ref{generalized distance}).

Both distances $\Delta$ and $\Delta_\MMVV$ share several features: they involve integrations of functionals of the kinetic and potential energies along on-shell trajectories; they reduce to the original moduli-space distance \cite{Ooguri:2006in} in the limit of a vanishing potential; they are positive.
Moreover, although physical solutions have a preferred direction to flow, both $\Delta$ and $\Delta_\MMVV$ can also be safely interpreted as symmetric in the endpoints.
Another point of contact is that both functions $\Delta$ and $\smash{\Delta_\MMVV}$ are not generally homogeneous on the fields due to the scalar potential, contrary to the moduli-space distance.

It is now interesting to focus on an important difference: while $\Delta_\MMVV$ depends on the total energy of the initial configuration, $\Delta$ does not.
This is because the prescription we gave instructs us to compute $\Delta$ specifically along the solution that is an attractor, while $\Delta_\MMVV$ is computed along \emph{a} solution.
Attractor solutions share the property of being independent of the initial conditions, as we verified for instance in the ubiquitous example of exponential potentials.
Integrating along attractor flows also leads $\Delta$ to satisfying the triangle inequality.
Again in the example of exponentials, one finds that $\Delta(\phi_0, \phi) = \Delta(\phi_0, \phi') + \Delta(\phi', \phi)$ for any three points $\phi_0$, $\phi'$ and $\phi$.
On the other hand, $\Delta_\MMVV$ does not satisfy the triangle inequality due to the dependence on the initial conditions.
Yet, we stress that our proposal is restricted only to points that are connected by an attractor.

To conclude, below we compare a couple of relevant examples to gain insight on further differences.
\begin{itemize}
    \item In terms of a (fake/pseudo-)superpotential $\tilde{W}$, we can write $\smash{\Delta = \sqrt{(D-1)/(D-2)} \, \int_{\mathrm{C}} \de \tau \, \ab \tilde{W} \ab}$, as in eq.~(\ref{generalized distance - superpotential formulation}).
    Theories with minimal SUSY in dimension $D=4$ can be formulated in terms of a K\"{a}hler potential $K$ and a (true) superpotential $W$.
    The SUSY algebra requires complex superfields $A$, which correspond to two real scalars.
    If $\epsilon=1$, we can make the identification $\smash{\tilde{W} = 2 \, \e^{K/2} \sqrt{W \overline{W}}}$.
    On the other hand, ref.~\cite{Mohseni:2024njl} computes distances between 4d SUSY solutions as
    \begin{equation}
        \Delta_\MMVV = \dfrac{1}{\sqrt{3}} \, \int_{\phi_i}^{\phi_e} \de \phi \, \dfrac{\ab \nabla W \ab}{\ab W \ab},
    \end{equation}
    where $\phi$ is a field such that $\smash{\dot{\phi}^2/2 = K_{A \overline{B}} \dot{A} \dot{\overline{B}}}$ along the on-shell trajectory, $\smash{\nabla_A W = (\der_A + K_A) W}$ and contractions are with respect to the K\"{a}hler metric $\smash{K_{A \overline{B}} = \der_A \der_{\overline{B}} K}$.
    Through eqs. (\ref{flow 1}, \ref{flow 2}), we may write
    \begin{equation}
        \Delta_\MMVV = \sqrt{\dfrac{2}{3}} \int_{\mathrm{C}} \de \tau \, \dfrac{\ab \der \tilde{W} \ab^2}{\ab \tilde{W} \ab},
    \end{equation}
    where the contraction is with respect to the metric $G_{ij}$ of the real-valued fields. This distance is completely different from $\Delta$.
    For the special case for BPS domain walls, a different but similar distance was proposed by ref. \cite{Basile:2023rvm}; see also refs.~\cite{Basile:2022zee, Basile:2022sda} for earlier related works.
    \item We discussed extensively FLRW- and DW-type solutions for exponential potentials $V = \Lambda \, \e^{- \gamma \varphi}$.
    For definiteness, let $\epsilon=-1$ and $\Lambda>0$, with $\gamma<\gamma_{\mathrm{c}}$.
    While our notion of distance is computed to be $\Delta = (\gamma_{\mathrm{c}}/\gamma) (\phi - \phi_0)$, as in eq.~(\ref{extended distance - cosmological solution for exponential potential}), ref.~\cite{Mohseni:2024njl} finds
    \begin{equation}
        \Delta_\MMVV = \dfrac{\gamma}{\gamma_{\mathrm{c}}} \, (\phi - \phi_0).
    \end{equation}
    The two values differ critically in the limit of zero steepness, being $\smash{\Delta \overset{\gamma \to 0}{\to} \infty}$ and $\smash{\Delta_\MMVV \overset{\gamma \to 0}{\to} 0}$ at fixed $\Delta \phi = \phi - \phi_0$. Note that, in terms of the time variable, the distances read $\smash{\Delta = (2 \gamma_{\mathrm{c}} / \gamma^2) \, \ln \, \tau/\tau_0}$ and $\smash{\Delta_\MMVV = (2/\gamma_{\mathrm{c}}) \, \ln \tau/\tau_0}$.
\end{itemize}

As a final difference to note, we emphasize that the definition of $\Delta$ does not apply to compute distances between two nearby vacua in that it needs attractors, contrary to the more general proposal for $\Delta_\MMVV$.

\bibliographystyle{apsrev4-1}
\bibliography{refs.bib}

\begin{thebibliography}{69}%
\makeatletter
\providecommand \@ifxundefined [1]{%
 \@ifx{#1\undefined}
}%
\providecommand \@ifnum [1]{%
 \ifnum #1\expandafter \@firstoftwo
 \else \expandafter \@secondoftwo
 \fi
}%
\providecommand \@ifx [1]{%
 \ifx #1\expandafter \@firstoftwo
 \else \expandafter \@secondoftwo
 \fi
}%
\providecommand \natexlab [1]{#1}%
\providecommand \enquote  [1]{``#1''}%
\providecommand \bibnamefont  [1]{#1}%
\providecommand \bibfnamefont [1]{#1}%
\providecommand \citenamefont [1]{#1}%
\providecommand \href@noop [0]{\@secondoftwo}%
\providecommand \href [0]{\begingroup \@sanitize@url \@href}%
\providecommand \@href[1]{\@@startlink{#1}\@@href}%
\providecommand \@@href[1]{\endgroup#1\@@endlink}%
\providecommand \@sanitize@url [0]{\catcode `\\12\catcode `\$12\catcode `\&12\catcode `\#12\catcode `\^12\catcode `\_12\catcode `\%12\relax}%
\providecommand \@@startlink[1]{}%
\providecommand \@@endlink[0]{}%
\providecommand \url  [0]{\begingroup\@sanitize@url \@url }%
\providecommand \@url [1]{\endgroup\@href {#1}{\urlprefix }}%
\providecommand \urlprefix  [0]{URL }%
\providecommand \Eprint [0]{\href }%
\providecommand \doibase [0]{http://dx.doi.org/}%
\providecommand \selectlanguage [0]{\@gobble}%
\providecommand \bibinfo  [0]{\@secondoftwo}%
\providecommand \bibfield  [0]{\@secondoftwo}%
\providecommand \translation [1]{[#1]}%
\providecommand \BibitemOpen [0]{}%
\providecommand \bibitemStop [0]{}%
\providecommand \bibitemNoStop [0]{.\EOS\space}%
\providecommand \EOS [0]{\spacefactor3000\relax}%
\providecommand \BibitemShut  [1]{\csname bibitem#1\endcsname}%
\let\auto@bib@innerbib\@empty
\bibitem [{\citenamefont {Ooguri}\ and\ \citenamefont {Vafa}(2007)}]{Ooguri:2006in}%
  \BibitemOpen
  \bibfield  {author} {\bibinfo {author} {\bibfnamefont {H.}~\bibnamefont {Ooguri}}\ and\ \bibinfo {author} {\bibfnamefont {C.}~\bibnamefont {Vafa}},\ }\href {\doibase 10.1016/j.nuclphysb.2006.10.033} {\bibfield  {journal} {\bibinfo  {journal} {Nucl. Phys. B}\ }\textbf {\bibinfo {volume} {766}},\ \bibinfo {pages} {21} (\bibinfo {year} {2007})},\ \Eprint {http://arxiv.org/abs/hep-th/0605264} {arXiv:hep-th/0605264} \BibitemShut {NoStop}%
\bibitem [{\citenamefont {Stout}(2021)}]{Stout:2021ubb}%
  \BibitemOpen
  \bibfield  {author} {\bibinfo {author} {\bibfnamefont {J.}~\bibnamefont {Stout}},\ }\href@noop {} {\  (\bibinfo {year} {2021})},\ \Eprint {http://arxiv.org/abs/2106.11313} {arXiv:2106.11313 [hep-th]} \BibitemShut {NoStop}%
\bibitem [{\citenamefont {Palti}(2019)}]{Palti:2019pca}%
  \BibitemOpen
  \bibfield  {author} {\bibinfo {author} {\bibfnamefont {E.}~\bibnamefont {Palti}},\ }\href {\doibase 10.1002/prop.201900037} {\bibfield  {journal} {\bibinfo  {journal} {Fortsch. Phys.}\ }\textbf {\bibinfo {volume} {67}},\ \bibinfo {pages} {1900037} (\bibinfo {year} {2019})},\ \Eprint {http://arxiv.org/abs/1903.06239} {arXiv:1903.06239 [hep-th]} \BibitemShut {NoStop}%
\bibitem [{\citenamefont {van Beest}\ \emph {et~al.}(2022)\citenamefont {van Beest}, \citenamefont {Calder\'on-Infante}, \citenamefont {Mirfendereski},\ and\ \citenamefont {Valenzuela}}]{vanBeest:2021lhn}%
  \BibitemOpen
  \bibfield  {author} {\bibinfo {author} {\bibfnamefont {M.}~\bibnamefont {van Beest}}, \bibinfo {author} {\bibfnamefont {J.}~\bibnamefont {Calder\'on-Infante}}, \bibinfo {author} {\bibfnamefont {D.}~\bibnamefont {Mirfendereski}}, \ and\ \bibinfo {author} {\bibfnamefont {I.}~\bibnamefont {Valenzuela}},\ }\href {\doibase 10.1016/j.physrep.2022.09.002} {\bibfield  {journal} {\bibinfo  {journal} {Phys. Rept.}\ }\textbf {\bibinfo {volume} {989}},\ \bibinfo {pages} {1} (\bibinfo {year} {2022})},\ \Eprint {http://arxiv.org/abs/2102.01111} {arXiv:2102.01111 [hep-th]} \BibitemShut {NoStop}%
\bibitem [{\citenamefont {Agmon}\ \emph {et~al.}(2022)\citenamefont {Agmon}, \citenamefont {Bedroya}, \citenamefont {Kang},\ and\ \citenamefont {Vafa}}]{Agmon:2022thq}%
  \BibitemOpen
  \bibfield  {author} {\bibinfo {author} {\bibfnamefont {N.~B.}\ \bibnamefont {Agmon}}, \bibinfo {author} {\bibfnamefont {A.}~\bibnamefont {Bedroya}}, \bibinfo {author} {\bibfnamefont {M.~J.}\ \bibnamefont {Kang}}, \ and\ \bibinfo {author} {\bibfnamefont {C.}~\bibnamefont {Vafa}},\ }\href@noop {} {\  (\bibinfo {year} {2022})},\ \Eprint {http://arxiv.org/abs/2212.06187} {arXiv:2212.06187 [hep-th]} \BibitemShut {NoStop}%
\bibitem [{\citenamefont {Lee}\ \emph {et~al.}(2022{\natexlab{a}})\citenamefont {Lee}, \citenamefont {Lerche},\ and\ \citenamefont {Weigand}}]{Lee:2019xtm}%
  \BibitemOpen
  \bibfield  {author} {\bibinfo {author} {\bibfnamefont {S.-J.}\ \bibnamefont {Lee}}, \bibinfo {author} {\bibfnamefont {W.}~\bibnamefont {Lerche}}, \ and\ \bibinfo {author} {\bibfnamefont {T.}~\bibnamefont {Weigand}},\ }\href {\doibase 10.1007/JHEP02(2022)096} {\bibfield  {journal} {\bibinfo  {journal} {JHEP}\ }\textbf {\bibinfo {volume} {02}},\ \bibinfo {pages} {096} (\bibinfo {year} {2022}{\natexlab{a}})},\ \Eprint {http://arxiv.org/abs/1904.06344} {arXiv:1904.06344 [hep-th]} \BibitemShut {NoStop}%
\bibitem [{\citenamefont {Lee}\ \emph {et~al.}(2022{\natexlab{b}})\citenamefont {Lee}, \citenamefont {Lerche},\ and\ \citenamefont {Weigand}}]{Lee:2019wij}%
  \BibitemOpen
  \bibfield  {author} {\bibinfo {author} {\bibfnamefont {S.-J.}\ \bibnamefont {Lee}}, \bibinfo {author} {\bibfnamefont {W.}~\bibnamefont {Lerche}}, \ and\ \bibinfo {author} {\bibfnamefont {T.}~\bibnamefont {Weigand}},\ }\href {\doibase 10.1007/JHEP02(2022)190} {\bibfield  {journal} {\bibinfo  {journal} {JHEP}\ }\textbf {\bibinfo {volume} {02}},\ \bibinfo {pages} {190} (\bibinfo {year} {2022}{\natexlab{b}})},\ \Eprint {http://arxiv.org/abs/1910.01135} {arXiv:1910.01135 [hep-th]} \BibitemShut {NoStop}%
\bibitem [{\citenamefont {Vafa}(2005)}]{Vafa:2005ui}%
  \BibitemOpen
  \bibfield  {author} {\bibinfo {author} {\bibfnamefont {C.}~\bibnamefont {Vafa}},\ }\href@noop {} {\  (\bibinfo {year} {2005})},\ \Eprint {http://arxiv.org/abs/hep-th/0509212} {arXiv:hep-th/0509212} \BibitemShut {NoStop}%
\bibitem [{\citenamefont {Schimmrigk}(2018)}]{Schimmrigk:2018gch}%
  \BibitemOpen
  \bibfield  {author} {\bibinfo {author} {\bibfnamefont {R.}~\bibnamefont {Schimmrigk}},\ }\href@noop {} {\  (\bibinfo {year} {2018})},\ \Eprint {http://arxiv.org/abs/1810.11699} {arXiv:1810.11699 [hep-th]} \BibitemShut {NoStop}%
\bibitem [{\citenamefont {Basile}\ and\ \citenamefont {Montella}(2024)}]{Basile:2023rvm}%
  \BibitemOpen
  \bibfield  {author} {\bibinfo {author} {\bibfnamefont {I.}~\bibnamefont {Basile}}\ and\ \bibinfo {author} {\bibfnamefont {C.}~\bibnamefont {Montella}},\ }\href {\doibase 10.1007/JHEP02(2024)227} {\bibfield  {journal} {\bibinfo  {journal} {JHEP}\ }\textbf {\bibinfo {volume} {02}},\ \bibinfo {pages} {227} (\bibinfo {year} {2024})},\ \Eprint {http://arxiv.org/abs/2309.04519} {arXiv:2309.04519 [hep-th]} \BibitemShut {NoStop}%
\bibitem [{\citenamefont {Basile}(2023)}]{Basile:2022zee}%
  \BibitemOpen
  \bibfield  {author} {\bibinfo {author} {\bibfnamefont {I.}~\bibnamefont {Basile}},\ }\href {\doibase 10.3390/astronomy2030015} {\bibfield  {journal} {\bibinfo  {journal} {Astronomy}\ }\textbf {\bibinfo {volume} {2}},\ \bibinfo {pages} {206} (\bibinfo {year} {2023})},\ \Eprint {http://arxiv.org/abs/2201.08851} {arXiv:2201.08851 [hep-th]} \BibitemShut {NoStop}%
\bibitem [{\citenamefont {Basile}\ \emph {et~al.}(2023)\citenamefont {Basile}, \citenamefont {Campoleoni}, \citenamefont {Pekar},\ and\ \citenamefont {Skvortsov}}]{Basile:2022sda}%
  \BibitemOpen
  \bibfield  {author} {\bibinfo {author} {\bibfnamefont {I.}~\bibnamefont {Basile}}, \bibinfo {author} {\bibfnamefont {A.}~\bibnamefont {Campoleoni}}, \bibinfo {author} {\bibfnamefont {S.}~\bibnamefont {Pekar}}, \ and\ \bibinfo {author} {\bibfnamefont {E.}~\bibnamefont {Skvortsov}},\ }\href {\doibase 10.1007/JHEP03(2023)075} {\bibfield  {journal} {\bibinfo  {journal} {JHEP}\ }\textbf {\bibinfo {volume} {03}},\ \bibinfo {pages} {075} (\bibinfo {year} {2023})},\ \Eprint {http://arxiv.org/abs/2209.14379} {arXiv:2209.14379 [hep-th]} \BibitemShut {NoStop}%
\bibitem [{\citenamefont {Mohseni}\ \emph {et~al.}(2024)\citenamefont {Mohseni}, \citenamefont {Montero}, \citenamefont {Vafa},\ and\ \citenamefont {Valenzuela}}]{Mohseni:2024njl}%
  \BibitemOpen
  \bibfield  {author} {\bibinfo {author} {\bibfnamefont {A.}~\bibnamefont {Mohseni}}, \bibinfo {author} {\bibfnamefont {M.}~\bibnamefont {Montero}}, \bibinfo {author} {\bibfnamefont {C.}~\bibnamefont {Vafa}}, \ and\ \bibinfo {author} {\bibfnamefont {I.}~\bibnamefont {Valenzuela}},\ }\href {\doibase 10.1007/JHEP12(2024)168} {\bibfield  {journal} {\bibinfo  {journal} {JHEP}\ }\textbf {\bibinfo {volume} {12}},\ \bibinfo {pages} {168} (\bibinfo {year} {2024})},\ \Eprint {http://arxiv.org/abs/2407.02705} {arXiv:2407.02705 [hep-th]} \BibitemShut {NoStop}%
\bibitem [{\citenamefont {Ooguri}\ \emph {et~al.}(2019)\citenamefont {Ooguri}, \citenamefont {Palti}, \citenamefont {Shiu},\ and\ \citenamefont {Vafa}}]{Ooguri:2018wrx}%
  \BibitemOpen
  \bibfield  {author} {\bibinfo {author} {\bibfnamefont {H.}~\bibnamefont {Ooguri}}, \bibinfo {author} {\bibfnamefont {E.}~\bibnamefont {Palti}}, \bibinfo {author} {\bibfnamefont {G.}~\bibnamefont {Shiu}}, \ and\ \bibinfo {author} {\bibfnamefont {C.}~\bibnamefont {Vafa}},\ }\href {\doibase 10.1016/j.physletb.2018.11.018} {\bibfield  {journal} {\bibinfo  {journal} {Phys. Lett. B}\ }\textbf {\bibinfo {volume} {788}},\ \bibinfo {pages} {180} (\bibinfo {year} {2019})},\ \Eprint {http://arxiv.org/abs/1810.05506} {arXiv:1810.05506 [hep-th]} \BibitemShut {NoStop}%
\bibitem [{\citenamefont {Hebecker}\ and\ \citenamefont {Wrase}(2019)}]{Hebecker:2018vxz}%
  \BibitemOpen
  \bibfield  {author} {\bibinfo {author} {\bibfnamefont {A.}~\bibnamefont {Hebecker}}\ and\ \bibinfo {author} {\bibfnamefont {T.}~\bibnamefont {Wrase}},\ }\href {\doibase 10.1002/prop.201800097} {\bibfield  {journal} {\bibinfo  {journal} {Fortsch. Phys.}\ }\textbf {\bibinfo {volume} {67}},\ \bibinfo {pages} {1800097} (\bibinfo {year} {2019})},\ \Eprint {http://arxiv.org/abs/1810.08182} {arXiv:1810.08182 [hep-th]} \BibitemShut {NoStop}%
\bibitem [{\citenamefont {Uzawa}(2018)}]{Uzawa:2018sal}%
  \BibitemOpen
  \bibfield  {author} {\bibinfo {author} {\bibfnamefont {K.}~\bibnamefont {Uzawa}},\ }\href {\doibase 10.1007/JHEP06(2018)041} {\bibfield  {journal} {\bibinfo  {journal} {JHEP}\ }\textbf {\bibinfo {volume} {06}},\ \bibinfo {pages} {041} (\bibinfo {year} {2018})},\ \Eprint {http://arxiv.org/abs/1803.11084} {arXiv:1803.11084 [hep-th]} \BibitemShut {NoStop}%
\bibitem [{\citenamefont {Bergshoeff}\ \emph {et~al.}(2009)\citenamefont {Bergshoeff}, \citenamefont {Chemissany}, \citenamefont {Ploegh}, \citenamefont {Trigiante},\ and\ \citenamefont {Van~Riet}}]{Bergshoeff:2008be}%
  \BibitemOpen
  \bibfield  {author} {\bibinfo {author} {\bibfnamefont {E.}~\bibnamefont {Bergshoeff}}, \bibinfo {author} {\bibfnamefont {W.}~\bibnamefont {Chemissany}}, \bibinfo {author} {\bibfnamefont {A.}~\bibnamefont {Ploegh}}, \bibinfo {author} {\bibfnamefont {M.}~\bibnamefont {Trigiante}}, \ and\ \bibinfo {author} {\bibfnamefont {T.}~\bibnamefont {Van~Riet}},\ }\href {\doibase 10.1016/j.nuclphysb.2008.10.023} {\bibfield  {journal} {\bibinfo  {journal} {Nucl. Phys. B}\ }\textbf {\bibinfo {volume} {812}},\ \bibinfo {pages} {343} (\bibinfo {year} {2009})},\ \Eprint {http://arxiv.org/abs/0806.2310} {arXiv:0806.2310 [hep-th]} \BibitemShut {NoStop}%
\bibitem [{\citenamefont {Freedman}\ \emph {et~al.}(2004)\citenamefont {Freedman}, \citenamefont {Nunez}, \citenamefont {Schnabl},\ and\ \citenamefont {Skenderis}}]{Freedman:2003ax}%
  \BibitemOpen
  \bibfield  {author} {\bibinfo {author} {\bibfnamefont {D.~Z.}\ \bibnamefont {Freedman}}, \bibinfo {author} {\bibfnamefont {C.}~\bibnamefont {Nunez}}, \bibinfo {author} {\bibfnamefont {M.}~\bibnamefont {Schnabl}}, \ and\ \bibinfo {author} {\bibfnamefont {K.}~\bibnamefont {Skenderis}},\ }\href {\doibase 10.1103/PhysRevD.69.104027} {\bibfield  {journal} {\bibinfo  {journal} {Phys. Rev. D}\ }\textbf {\bibinfo {volume} {69}},\ \bibinfo {pages} {104027} (\bibinfo {year} {2004})},\ \Eprint {http://arxiv.org/abs/hep-th/0312055} {arXiv:hep-th/0312055} \BibitemShut {NoStop}%
\bibitem [{\citenamefont {Skenderis}\ and\ \citenamefont {Townsend}(2006{\natexlab{a}})}]{Skenderis:2006jq}%
  \BibitemOpen
  \bibfield  {author} {\bibinfo {author} {\bibfnamefont {K.}~\bibnamefont {Skenderis}}\ and\ \bibinfo {author} {\bibfnamefont {P.~K.}\ \bibnamefont {Townsend}},\ }\href {\doibase 10.1103/PhysRevLett.96.191301} {\bibfield  {journal} {\bibinfo  {journal} {Phys. Rev. Lett.}\ }\textbf {\bibinfo {volume} {96}},\ \bibinfo {pages} {191301} (\bibinfo {year} {2006}{\natexlab{a}})},\ \Eprint {http://arxiv.org/abs/hep-th/0602260} {arXiv:hep-th/0602260} \BibitemShut {NoStop}%
\bibitem [{\citenamefont {Skenderis}\ and\ \citenamefont {Townsend}(2006{\natexlab{b}})}]{Skenderis:2006rr}%
  \BibitemOpen
  \bibfield  {author} {\bibinfo {author} {\bibfnamefont {K.}~\bibnamefont {Skenderis}}\ and\ \bibinfo {author} {\bibfnamefont {P.~K.}\ \bibnamefont {Townsend}},\ }\href {\doibase 10.1103/PhysRevD.74.125008} {\bibfield  {journal} {\bibinfo  {journal} {Phys. Rev. D}\ }\textbf {\bibinfo {volume} {74}},\ \bibinfo {pages} {125008} (\bibinfo {year} {2006}{\natexlab{b}})},\ \Eprint {http://arxiv.org/abs/hep-th/0609056} {arXiv:hep-th/0609056} \BibitemShut {NoStop}%
\bibitem [{\citenamefont {de~Boer}\ \emph {et~al.}(2000)\citenamefont {de~Boer}, \citenamefont {Verlinde},\ and\ \citenamefont {Verlinde}}]{deBoer:1999tgo}%
  \BibitemOpen
  \bibfield  {author} {\bibinfo {author} {\bibfnamefont {J.}~\bibnamefont {de~Boer}}, \bibinfo {author} {\bibfnamefont {E.~P.}\ \bibnamefont {Verlinde}}, \ and\ \bibinfo {author} {\bibfnamefont {H.~L.}\ \bibnamefont {Verlinde}},\ }\href {\doibase 10.1088/1126-6708/2000/08/003} {\bibfield  {journal} {\bibinfo  {journal} {JHEP}\ }\textbf {\bibinfo {volume} {08}},\ \bibinfo {pages} {003} (\bibinfo {year} {2000})},\ \Eprint {http://arxiv.org/abs/hep-th/9912012} {arXiv:hep-th/9912012} \BibitemShut {NoStop}%
\bibitem [{\citenamefont {Diaz~Dorronsoro}\ \emph {et~al.}(2017)\citenamefont {Diaz~Dorronsoro}, \citenamefont {Truijen},\ and\ \citenamefont {Van~Riet}}]{DiazDorronsoro:2016rrz}%
  \BibitemOpen
  \bibfield  {author} {\bibinfo {author} {\bibfnamefont {J.}~\bibnamefont {Diaz~Dorronsoro}}, \bibinfo {author} {\bibfnamefont {B.}~\bibnamefont {Truijen}}, \ and\ \bibinfo {author} {\bibfnamefont {T.}~\bibnamefont {Van~Riet}},\ }\href {\doibase 10.1088/1361-6382/aa64b4} {\bibfield  {journal} {\bibinfo  {journal} {Class. Quant. Grav.}\ }\textbf {\bibinfo {volume} {34}},\ \bibinfo {pages} {095003} (\bibinfo {year} {2017})},\ \Eprint {http://arxiv.org/abs/1606.07730} {arXiv:1606.07730 [hep-th]} \BibitemShut {NoStop}%
\bibitem [{\citenamefont {Chemissany}\ \emph {et~al.}(2007)\citenamefont {Chemissany}, \citenamefont {Ploegh},\ and\ \citenamefont {Van~Riet}}]{Chemissany:2007fg}%
  \BibitemOpen
  \bibfield  {author} {\bibinfo {author} {\bibfnamefont {W.}~\bibnamefont {Chemissany}}, \bibinfo {author} {\bibfnamefont {A.}~\bibnamefont {Ploegh}}, \ and\ \bibinfo {author} {\bibfnamefont {T.}~\bibnamefont {Van~Riet}},\ }\href {\doibase 10.1088/0264-9381/24/18/009} {\bibfield  {journal} {\bibinfo  {journal} {Class. Quant. Grav.}\ }\textbf {\bibinfo {volume} {24}},\ \bibinfo {pages} {4679} (\bibinfo {year} {2007})},\ \Eprint {http://arxiv.org/abs/0704.1653} {arXiv:0704.1653 [hep-th]} \BibitemShut {NoStop}%
\bibitem [{\citenamefont {Copeland}\ \emph {et~al.}(1998)\citenamefont {Copeland}, \citenamefont {Liddle},\ and\ \citenamefont {Wands}}]{Copeland:1997et}%
  \BibitemOpen
  \bibfield  {author} {\bibinfo {author} {\bibfnamefont {E.~J.}\ \bibnamefont {Copeland}}, \bibinfo {author} {\bibfnamefont {A.~R.}\ \bibnamefont {Liddle}}, \ and\ \bibinfo {author} {\bibfnamefont {D.}~\bibnamefont {Wands}},\ }\href {\doibase 10.1103/PhysRevD.57.4686} {\bibfield  {journal} {\bibinfo  {journal} {Phys. Rev. D}\ }\textbf {\bibinfo {volume} {57}},\ \bibinfo {pages} {4686} (\bibinfo {year} {1998})},\ \Eprint {http://arxiv.org/abs/gr-qc/9711068} {arXiv:gr-qc/9711068} \BibitemShut {NoStop}%
\bibitem [{\citenamefont {Etheredge}\ \emph {et~al.}(2022)\citenamefont {Etheredge}, \citenamefont {Heidenreich}, \citenamefont {Kaya}, \citenamefont {Qiu},\ and\ \citenamefont {Rudelius}}]{Etheredge:2022opl}%
  \BibitemOpen
  \bibfield  {author} {\bibinfo {author} {\bibfnamefont {M.}~\bibnamefont {Etheredge}}, \bibinfo {author} {\bibfnamefont {B.}~\bibnamefont {Heidenreich}}, \bibinfo {author} {\bibfnamefont {S.}~\bibnamefont {Kaya}}, \bibinfo {author} {\bibfnamefont {Y.}~\bibnamefont {Qiu}}, \ and\ \bibinfo {author} {\bibfnamefont {T.}~\bibnamefont {Rudelius}},\ }\href {\doibase 10.1007/JHEP12(2022)114} {\bibfield  {journal} {\bibinfo  {journal} {JHEP}\ }\textbf {\bibinfo {volume} {12}},\ \bibinfo {pages} {114} (\bibinfo {year} {2022})},\ \Eprint {http://arxiv.org/abs/2206.04063} {arXiv:2206.04063 [hep-th]} \BibitemShut {NoStop}%
\bibitem [{\citenamefont {Van~Riet}\ and\ \citenamefont {Zoccarato}(2024)}]{VanRiet:2023pnx}%
  \BibitemOpen
  \bibfield  {author} {\bibinfo {author} {\bibfnamefont {T.}~\bibnamefont {Van~Riet}}\ and\ \bibinfo {author} {\bibfnamefont {G.}~\bibnamefont {Zoccarato}},\ }\href {\doibase 10.1016/j.physrep.2023.11.003} {\bibfield  {journal} {\bibinfo  {journal} {Phys. Rept.}\ }\textbf {\bibinfo {volume} {1049}},\ \bibinfo {pages} {1} (\bibinfo {year} {2024})},\ \Eprint {http://arxiv.org/abs/2305.01722} {arXiv:2305.01722 [hep-th]} \BibitemShut {NoStop}%
\bibitem [{\citenamefont {Shiu}\ \emph {et~al.}(2023{\natexlab{a}})\citenamefont {Shiu}, \citenamefont {Tonioni},\ and\ \citenamefont {Tran}}]{Shiu:2023fhb}%
  \BibitemOpen
  \bibfield  {author} {\bibinfo {author} {\bibfnamefont {G.}~\bibnamefont {Shiu}}, \bibinfo {author} {\bibfnamefont {F.}~\bibnamefont {Tonioni}}, \ and\ \bibinfo {author} {\bibfnamefont {H.~V.}\ \bibnamefont {Tran}},\ }\href {\doibase 10.1103/PhysRevD.108.063528} {\bibfield  {journal} {\bibinfo  {journal} {Phys. Rev. D}\ }\textbf {\bibinfo {volume} {108}},\ \bibinfo {pages} {063528} (\bibinfo {year} {2023}{\natexlab{a}})},\ \Eprint {http://arxiv.org/abs/2306.07327} {arXiv:2306.07327 [hep-th]} \BibitemShut {NoStop}%
\bibitem [{\citenamefont {Sonner}\ and\ \citenamefont {Townsend}(2006{\natexlab{a}})}]{Sonner:2006yn}%
  \BibitemOpen
  \bibfield  {author} {\bibinfo {author} {\bibfnamefont {J.}~\bibnamefont {Sonner}}\ and\ \bibinfo {author} {\bibfnamefont {P.~K.}\ \bibnamefont {Townsend}},\ }\href {\doibase 10.1103/PhysRevD.74.103508} {\bibfield  {journal} {\bibinfo  {journal} {Phys. Rev. D}\ }\textbf {\bibinfo {volume} {74}},\ \bibinfo {pages} {103508} (\bibinfo {year} {2006}{\natexlab{a}})},\ \Eprint {http://arxiv.org/abs/hep-th/0608068} {arXiv:hep-th/0608068} \BibitemShut {NoStop}%
\bibitem [{\citenamefont {Cicoli}\ \emph {et~al.}(2020{\natexlab{a}})\citenamefont {Cicoli}, \citenamefont {Dibitetto},\ and\ \citenamefont {Pedro}}]{Cicoli:2020cfj}%
  \BibitemOpen
  \bibfield  {author} {\bibinfo {author} {\bibfnamefont {M.}~\bibnamefont {Cicoli}}, \bibinfo {author} {\bibfnamefont {G.}~\bibnamefont {Dibitetto}}, \ and\ \bibinfo {author} {\bibfnamefont {F.~G.}\ \bibnamefont {Pedro}},\ }\href {\doibase 10.1103/PhysRevD.101.103524} {\bibfield  {journal} {\bibinfo  {journal} {Phys. Rev. D}\ }\textbf {\bibinfo {volume} {101}},\ \bibinfo {pages} {103524} (\bibinfo {year} {2020}{\natexlab{a}})},\ \Eprint {http://arxiv.org/abs/2002.02695} {arXiv:2002.02695 [gr-qc]} \BibitemShut {NoStop}%
\bibitem [{\citenamefont {Revello}(2024)}]{Revello:2023hro}%
  \BibitemOpen
  \bibfield  {author} {\bibinfo {author} {\bibfnamefont {F.}~\bibnamefont {Revello}},\ }\href {\doibase 10.1007/JHEP05(2024)037} {\bibfield  {journal} {\bibinfo  {journal} {JHEP}\ }\textbf {\bibinfo {volume} {05}},\ \bibinfo {pages} {037} (\bibinfo {year} {2024})},\ \Eprint {http://arxiv.org/abs/2311.12429} {arXiv:2311.12429 [hep-th]} \BibitemShut {NoStop}%
\bibitem [{\citenamefont {Shiu}\ \emph {et~al.}(2024{\natexlab{a}})\citenamefont {Shiu}, \citenamefont {Tonioni},\ and\ \citenamefont {Tran}}]{Shiu:2024sbe}%
  \BibitemOpen
  \bibfield  {author} {\bibinfo {author} {\bibfnamefont {G.}~\bibnamefont {Shiu}}, \bibinfo {author} {\bibfnamefont {F.}~\bibnamefont {Tonioni}}, \ and\ \bibinfo {author} {\bibfnamefont {H.~V.}\ \bibnamefont {Tran}},\ }\href {\doibase 10.1007/JHEP09(2024)158} {\bibfield  {journal} {\bibinfo  {journal} {JHEP}\ }\textbf {\bibinfo {volume} {09}},\ \bibinfo {pages} {158} (\bibinfo {year} {2024}{\natexlab{a}})},\ \Eprint {http://arxiv.org/abs/2406.17030} {arXiv:2406.17030 [hep-th]} \BibitemShut {NoStop}%
\bibitem [{\citenamefont {Collinucci}\ \emph {et~al.}(2005)\citenamefont {Collinucci}, \citenamefont {Nielsen},\ and\ \citenamefont {Van~Riet}}]{Collinucci:2004iw}%
  \BibitemOpen
  \bibfield  {author} {\bibinfo {author} {\bibfnamefont {A.}~\bibnamefont {Collinucci}}, \bibinfo {author} {\bibfnamefont {M.}~\bibnamefont {Nielsen}}, \ and\ \bibinfo {author} {\bibfnamefont {T.}~\bibnamefont {Van~Riet}},\ }\href {\doibase 10.1088/0264-9381/22/7/005} {\bibfield  {journal} {\bibinfo  {journal} {Class. Quant. Grav.}\ }\textbf {\bibinfo {volume} {22}},\ \bibinfo {pages} {1269} (\bibinfo {year} {2005})},\ \Eprint {http://arxiv.org/abs/hep-th/0407047} {arXiv:hep-th/0407047} \BibitemShut {NoStop}%
\bibitem [{\citenamefont {Hartong}\ \emph {et~al.}(2006)\citenamefont {Hartong}, \citenamefont {Ploegh}, \citenamefont {Van~Riet},\ and\ \citenamefont {Westra}}]{Hartong:2006rt}%
  \BibitemOpen
  \bibfield  {author} {\bibinfo {author} {\bibfnamefont {J.}~\bibnamefont {Hartong}}, \bibinfo {author} {\bibfnamefont {A.}~\bibnamefont {Ploegh}}, \bibinfo {author} {\bibfnamefont {T.}~\bibnamefont {Van~Riet}}, \ and\ \bibinfo {author} {\bibfnamefont {D.~B.}\ \bibnamefont {Westra}},\ }\href {\doibase 10.1088/0264-9381/23/14/003} {\bibfield  {journal} {\bibinfo  {journal} {Class. Quant. Grav.}\ }\textbf {\bibinfo {volume} {23}},\ \bibinfo {pages} {4593} (\bibinfo {year} {2006})},\ \Eprint {http://arxiv.org/abs/gr-qc/0602077} {arXiv:gr-qc/0602077} \BibitemShut {NoStop}%
\bibitem [{\citenamefont {Sonner}\ and\ \citenamefont {Townsend}(2007)}]{Sonner:2007cp}%
  \BibitemOpen
  \bibfield  {author} {\bibinfo {author} {\bibfnamefont {J.}~\bibnamefont {Sonner}}\ and\ \bibinfo {author} {\bibfnamefont {P.~K.}\ \bibnamefont {Townsend}},\ }\href {\doibase 10.1088/0264-9381/24/13/021} {\bibfield  {journal} {\bibinfo  {journal} {Class. Quant. Grav.}\ }\textbf {\bibinfo {volume} {24}},\ \bibinfo {pages} {3479} (\bibinfo {year} {2007})},\ \Eprint {http://arxiv.org/abs/hep-th/0703276} {arXiv:hep-th/0703276} \BibitemShut {NoStop}%
\bibitem [{\citenamefont {Karthauser}\ and\ \citenamefont {Saffin}(2006)}]{Karthauser:2006ix}%
  \BibitemOpen
  \bibfield  {author} {\bibinfo {author} {\bibfnamefont {J.~L.~P.}\ \bibnamefont {Karthauser}}\ and\ \bibinfo {author} {\bibfnamefont {P.~M.}\ \bibnamefont {Saffin}},\ }\href {\doibase 10.1088/0264-9381/23/14/004} {\bibfield  {journal} {\bibinfo  {journal} {Class. Quant. Grav.}\ }\textbf {\bibinfo {volume} {23}},\ \bibinfo {pages} {4615} (\bibinfo {year} {2006})},\ \Eprint {http://arxiv.org/abs/hep-th/0604046} {arXiv:hep-th/0604046} \BibitemShut {NoStop}%
\bibitem [{\citenamefont {Etheredge}\ and\ \citenamefont {Heidenreich}(2023)}]{Etheredge:2023zjk}%
  \BibitemOpen
  \bibfield  {author} {\bibinfo {author} {\bibfnamefont {M.}~\bibnamefont {Etheredge}}\ and\ \bibinfo {author} {\bibfnamefont {B.}~\bibnamefont {Heidenreich}},\ }\href@noop {} {\  (\bibinfo {year} {2023})},\ \Eprint {http://arxiv.org/abs/2311.18693} {arXiv:2311.18693 [hep-th]} \BibitemShut {NoStop}%
\bibitem [{\citenamefont {Sonner}\ and\ \citenamefont {Townsend}(2006{\natexlab{b}})}]{Sonner:2005sj}%
  \BibitemOpen
  \bibfield  {author} {\bibinfo {author} {\bibfnamefont {J.}~\bibnamefont {Sonner}}\ and\ \bibinfo {author} {\bibfnamefont {P.~K.}\ \bibnamefont {Townsend}},\ }\href {\doibase 10.1088/0264-9381/23/2/010} {\bibfield  {journal} {\bibinfo  {journal} {Class. Quant. Grav.}\ }\textbf {\bibinfo {volume} {23}},\ \bibinfo {pages} {441} (\bibinfo {year} {2006}{\natexlab{b}})},\ \Eprint {http://arxiv.org/abs/hep-th/0510115} {arXiv:hep-th/0510115} \BibitemShut {NoStop}%
\bibitem [{\citenamefont {Shiu}\ \emph {et~al.}(2023{\natexlab{b}})\citenamefont {Shiu}, \citenamefont {Tonioni},\ and\ \citenamefont {Tran}}]{Shiu:2023nph}%
  \BibitemOpen
  \bibfield  {author} {\bibinfo {author} {\bibfnamefont {G.}~\bibnamefont {Shiu}}, \bibinfo {author} {\bibfnamefont {F.}~\bibnamefont {Tonioni}}, \ and\ \bibinfo {author} {\bibfnamefont {H.~V.}\ \bibnamefont {Tran}},\ }\href {\doibase 10.1103/PhysRevD.108.063527} {\bibfield  {journal} {\bibinfo  {journal} {Phys. Rev. D}\ }\textbf {\bibinfo {volume} {108}},\ \bibinfo {pages} {063527} (\bibinfo {year} {2023}{\natexlab{b}})},\ \Eprint {http://arxiv.org/abs/2303.03418} {arXiv:2303.03418 [hep-th]} \BibitemShut {NoStop}%
\bibitem [{\citenamefont {Shiu}\ \emph {et~al.}(2024{\natexlab{b}})\citenamefont {Shiu}, \citenamefont {Tonioni},\ and\ \citenamefont {Tran}}]{Shiu:2023yzt}%
  \BibitemOpen
  \bibfield  {author} {\bibinfo {author} {\bibfnamefont {G.}~\bibnamefont {Shiu}}, \bibinfo {author} {\bibfnamefont {F.}~\bibnamefont {Tonioni}}, \ and\ \bibinfo {author} {\bibfnamefont {H.~V.}\ \bibnamefont {Tran}},\ }\href {\doibase 10.1088/1475-7516/2024/05/124} {\bibfield  {journal} {\bibinfo  {journal} {JCAP}\ }\textbf {\bibinfo {volume} {05}},\ \bibinfo {pages} {124} (\bibinfo {year} {2024}{\natexlab{b}})},\ \Eprint {http://arxiv.org/abs/2312.06772} {arXiv:2312.06772 [gr-qc]} \BibitemShut {NoStop}%
\bibitem [{\citenamefont {Cicoli}\ \emph {et~al.}(2020{\natexlab{b}})\citenamefont {Cicoli}, \citenamefont {Dibitetto},\ and\ \citenamefont {Pedro}}]{Cicoli:2020noz}%
  \BibitemOpen
  \bibfield  {author} {\bibinfo {author} {\bibfnamefont {M.}~\bibnamefont {Cicoli}}, \bibinfo {author} {\bibfnamefont {G.}~\bibnamefont {Dibitetto}}, \ and\ \bibinfo {author} {\bibfnamefont {F.~G.}\ \bibnamefont {Pedro}},\ }\href {\doibase 10.1007/JHEP10(2020)035} {\bibfield  {journal} {\bibinfo  {journal} {JHEP}\ }\textbf {\bibinfo {volume} {10}},\ \bibinfo {pages} {035} (\bibinfo {year} {2020}{\natexlab{b}})},\ \Eprint {http://arxiv.org/abs/2007.11011} {arXiv:2007.11011 [hep-th]} \BibitemShut {NoStop}%
\bibitem [{\citenamefont {Buratti}\ \emph {et~al.}(2019)\citenamefont {Buratti}, \citenamefont {Calder\'on},\ and\ \citenamefont {Uranga}}]{Buratti:2018xjt}%
  \BibitemOpen
  \bibfield  {author} {\bibinfo {author} {\bibfnamefont {G.}~\bibnamefont {Buratti}}, \bibinfo {author} {\bibfnamefont {J.}~\bibnamefont {Calder\'on}}, \ and\ \bibinfo {author} {\bibfnamefont {A.~M.}\ \bibnamefont {Uranga}},\ }\href {\doibase 10.1007/JHEP05(2019)176} {\bibfield  {journal} {\bibinfo  {journal} {JHEP}\ }\textbf {\bibinfo {volume} {05}},\ \bibinfo {pages} {176} (\bibinfo {year} {2019})},\ \Eprint {http://arxiv.org/abs/1812.05016} {arXiv:1812.05016 [hep-th]} \BibitemShut {NoStop}%
\bibitem [{\citenamefont {Klebanov}\ and\ \citenamefont {Strassler}(2000)}]{Klebanov:2000hb}%
  \BibitemOpen
  \bibfield  {author} {\bibinfo {author} {\bibfnamefont {I.~R.}\ \bibnamefont {Klebanov}}\ and\ \bibinfo {author} {\bibfnamefont {M.~J.}\ \bibnamefont {Strassler}},\ }\href {\doibase 10.1088/1126-6708/2000/08/052} {\bibfield  {journal} {\bibinfo  {journal} {JHEP}\ }\textbf {\bibinfo {volume} {08}},\ \bibinfo {pages} {052} (\bibinfo {year} {2000})},\ \Eprint {http://arxiv.org/abs/hep-th/0007191} {arXiv:hep-th/0007191} \BibitemShut {NoStop}%
\bibitem [{\citenamefont {Heidenreich}\ \emph {et~al.}(2016)\citenamefont {Heidenreich}, \citenamefont {Reece},\ and\ \citenamefont {Rudelius}}]{Heidenreich:2015nta}%
  \BibitemOpen
  \bibfield  {author} {\bibinfo {author} {\bibfnamefont {B.}~\bibnamefont {Heidenreich}}, \bibinfo {author} {\bibfnamefont {M.}~\bibnamefont {Reece}}, \ and\ \bibinfo {author} {\bibfnamefont {T.}~\bibnamefont {Rudelius}},\ }\href {\doibase 10.1007/JHEP02(2016)140} {\bibfield  {journal} {\bibinfo  {journal} {JHEP}\ }\textbf {\bibinfo {volume} {02}},\ \bibinfo {pages} {140} (\bibinfo {year} {2016})},\ \Eprint {http://arxiv.org/abs/1509.06374} {arXiv:1509.06374 [hep-th]} \BibitemShut {NoStop}%
\bibitem [{\citenamefont {Heidenreich}\ \emph {et~al.}(2019)\citenamefont {Heidenreich}, \citenamefont {Reece},\ and\ \citenamefont {Rudelius}}]{Heidenreich:2019zkl}%
  \BibitemOpen
  \bibfield  {author} {\bibinfo {author} {\bibfnamefont {B.}~\bibnamefont {Heidenreich}}, \bibinfo {author} {\bibfnamefont {M.}~\bibnamefont {Reece}}, \ and\ \bibinfo {author} {\bibfnamefont {T.}~\bibnamefont {Rudelius}},\ }\href {\doibase 10.1007/JHEP10(2019)055} {\bibfield  {journal} {\bibinfo  {journal} {JHEP}\ }\textbf {\bibinfo {volume} {10}},\ \bibinfo {pages} {055} (\bibinfo {year} {2019})},\ \Eprint {http://arxiv.org/abs/1906.02206} {arXiv:1906.02206 [hep-th]} \BibitemShut {NoStop}%
\bibitem [{\citenamefont {Rudelius}(2021{\natexlab{a}})}]{Rudelius:2021oaz}%
  \BibitemOpen
  \bibfield  {author} {\bibinfo {author} {\bibfnamefont {T.}~\bibnamefont {Rudelius}},\ }\href {\doibase 10.1007/JHEP08(2021)041} {\bibfield  {journal} {\bibinfo  {journal} {JHEP}\ }\textbf {\bibinfo {volume} {08}},\ \bibinfo {pages} {041} (\bibinfo {year} {2021}{\natexlab{a}})},\ \Eprint {http://arxiv.org/abs/2101.11617} {arXiv:2101.11617 [hep-th]} \BibitemShut {NoStop}%
\bibitem [{\citenamefont {van~de Heisteeg}\ \emph {et~al.}(2024{\natexlab{a}})\citenamefont {van~de Heisteeg}, \citenamefont {Vafa}, \citenamefont {Wiesner},\ and\ \citenamefont {Wu}}]{vandeHeisteeg:2023dlw}%
  \BibitemOpen
  \bibfield  {author} {\bibinfo {author} {\bibfnamefont {D.}~\bibnamefont {van~de Heisteeg}}, \bibinfo {author} {\bibfnamefont {C.}~\bibnamefont {Vafa}}, \bibinfo {author} {\bibfnamefont {M.}~\bibnamefont {Wiesner}}, \ and\ \bibinfo {author} {\bibfnamefont {D.~H.}\ \bibnamefont {Wu}},\ }\href {\doibase 10.1007/JHEP05(2024)112} {\bibfield  {journal} {\bibinfo  {journal} {JHEP}\ }\textbf {\bibinfo {volume} {05}},\ \bibinfo {pages} {112} (\bibinfo {year} {2024}{\natexlab{a}})},\ \Eprint {http://arxiv.org/abs/2310.07213} {arXiv:2310.07213 [hep-th]} \BibitemShut {NoStop}%
\bibitem [{\citenamefont {Rudelius}(2021{\natexlab{b}})}]{Rudelius:2021azq}%
  \BibitemOpen
  \bibfield  {author} {\bibinfo {author} {\bibfnamefont {T.}~\bibnamefont {Rudelius}},\ }\href {\doibase 10.1103/PhysRevD.104.126023} {\bibfield  {journal} {\bibinfo  {journal} {Phys. Rev. D}\ }\textbf {\bibinfo {volume} {104}},\ \bibinfo {pages} {126023} (\bibinfo {year} {2021}{\natexlab{b}})},\ \Eprint {http://arxiv.org/abs/2106.09026} {arXiv:2106.09026 [hep-th]} \BibitemShut {NoStop}%
\bibitem [{\citenamefont {Hebecker}\ \emph {et~al.}(2023)\citenamefont {Hebecker}, \citenamefont {Schreyer},\ and\ \citenamefont {Venken}}]{Hebecker:2023qke}%
  \BibitemOpen
  \bibfield  {author} {\bibinfo {author} {\bibfnamefont {A.}~\bibnamefont {Hebecker}}, \bibinfo {author} {\bibfnamefont {S.}~\bibnamefont {Schreyer}}, \ and\ \bibinfo {author} {\bibfnamefont {G.}~\bibnamefont {Venken}},\ }\href {\doibase 10.1007/JHEP11(2023)173} {\bibfield  {journal} {\bibinfo  {journal} {JHEP}\ }\textbf {\bibinfo {volume} {11}},\ \bibinfo {pages} {173} (\bibinfo {year} {2023})},\ \Eprint {http://arxiv.org/abs/2306.17213} {arXiv:2306.17213 [hep-th]} \BibitemShut {NoStop}%
\bibitem [{\citenamefont {Obied}\ \emph {et~al.}(2018)\citenamefont {Obied}, \citenamefont {Ooguri}, \citenamefont {Spodyneiko},\ and\ \citenamefont {Vafa}}]{Obied:2018sgi}%
  \BibitemOpen
  \bibfield  {author} {\bibinfo {author} {\bibfnamefont {G.}~\bibnamefont {Obied}}, \bibinfo {author} {\bibfnamefont {H.}~\bibnamefont {Ooguri}}, \bibinfo {author} {\bibfnamefont {L.}~\bibnamefont {Spodyneiko}}, \ and\ \bibinfo {author} {\bibfnamefont {C.}~\bibnamefont {Vafa}},\ }\href@noop {} {\  (\bibinfo {year} {2018})},\ \Eprint {http://arxiv.org/abs/1806.08362} {arXiv:1806.08362 [hep-th]} \BibitemShut {NoStop}%
\bibitem [{\citenamefont {Bernardo}\ and\ \citenamefont {Brandenberger}(2021)}]{Bernardo:2021wnv}%
  \BibitemOpen
  \bibfield  {author} {\bibinfo {author} {\bibfnamefont {H.}~\bibnamefont {Bernardo}}\ and\ \bibinfo {author} {\bibfnamefont {R.}~\bibnamefont {Brandenberger}},\ }\href {\doibase 10.1007/JHEP07(2021)206} {\bibfield  {journal} {\bibinfo  {journal} {JHEP}\ }\textbf {\bibinfo {volume} {07}},\ \bibinfo {pages} {206} (\bibinfo {year} {2021})},\ \Eprint {http://arxiv.org/abs/2104.00630} {arXiv:2104.00630 [hep-th]} \BibitemShut {NoStop}%
\bibitem [{\citenamefont {Andriot}\ \emph {et~al.}(2023)\citenamefont {Andriot}, \citenamefont {Horer},\ and\ \citenamefont {Tringas}}]{Andriot:2022brg}%
  \BibitemOpen
  \bibfield  {author} {\bibinfo {author} {\bibfnamefont {D.}~\bibnamefont {Andriot}}, \bibinfo {author} {\bibfnamefont {L.}~\bibnamefont {Horer}}, \ and\ \bibinfo {author} {\bibfnamefont {G.}~\bibnamefont {Tringas}},\ }\href {\doibase 10.1007/JHEP04(2023)139} {\bibfield  {journal} {\bibinfo  {journal} {JHEP}\ }\textbf {\bibinfo {volume} {04}},\ \bibinfo {pages} {139} (\bibinfo {year} {2023})},\ \Eprint {http://arxiv.org/abs/2212.04517} {arXiv:2212.04517 [hep-th]} \BibitemShut {NoStop}%
\bibitem [{\citenamefont {Khoury}\ \emph {et~al.}(2001)\citenamefont {Khoury}, \citenamefont {Ovrut}, \citenamefont {Steinhardt},\ and\ \citenamefont {Turok}}]{Khoury:2001wf}%
  \BibitemOpen
  \bibfield  {author} {\bibinfo {author} {\bibfnamefont {J.}~\bibnamefont {Khoury}}, \bibinfo {author} {\bibfnamefont {B.~A.}\ \bibnamefont {Ovrut}}, \bibinfo {author} {\bibfnamefont {P.~J.}\ \bibnamefont {Steinhardt}}, \ and\ \bibinfo {author} {\bibfnamefont {N.}~\bibnamefont {Turok}},\ }\href {\doibase 10.1103/PhysRevD.64.123522} {\bibfield  {journal} {\bibinfo  {journal} {Phys. Rev. D}\ }\textbf {\bibinfo {volume} {64}},\ \bibinfo {pages} {123522} (\bibinfo {year} {2001})},\ \Eprint {http://arxiv.org/abs/hep-th/0103239} {arXiv:hep-th/0103239} \BibitemShut {NoStop}%
\bibitem [{\citenamefont {Khoury}\ \emph {et~al.}(2002)\citenamefont {Khoury}, \citenamefont {Ovrut}, \citenamefont {Seiberg}, \citenamefont {Steinhardt},\ and\ \citenamefont {Turok}}]{Khoury:2001bz}%
  \BibitemOpen
  \bibfield  {author} {\bibinfo {author} {\bibfnamefont {J.}~\bibnamefont {Khoury}}, \bibinfo {author} {\bibfnamefont {B.~A.}\ \bibnamefont {Ovrut}}, \bibinfo {author} {\bibfnamefont {N.}~\bibnamefont {Seiberg}}, \bibinfo {author} {\bibfnamefont {P.~J.}\ \bibnamefont {Steinhardt}}, \ and\ \bibinfo {author} {\bibfnamefont {N.}~\bibnamefont {Turok}},\ }\href {\doibase 10.1103/PhysRevD.65.086007} {\bibfield  {journal} {\bibinfo  {journal} {Phys. Rev. D}\ }\textbf {\bibinfo {volume} {65}},\ \bibinfo {pages} {086007} (\bibinfo {year} {2002})},\ \Eprint {http://arxiv.org/abs/hep-th/0108187} {arXiv:hep-th/0108187} \BibitemShut {NoStop}%
\bibitem [{\citenamefont {Steinhardt}\ \emph {et~al.}(2002)\citenamefont {Steinhardt}, \citenamefont {Turok},\ and\ \citenamefont {Turok}}]{Steinhardt:2002ih}%
  \BibitemOpen
  \bibfield  {author} {\bibinfo {author} {\bibfnamefont {P.~J.}\ \bibnamefont {Steinhardt}}, \bibinfo {author} {\bibfnamefont {N.}~\bibnamefont {Turok}}, \ and\ \bibinfo {author} {\bibfnamefont {N.}~\bibnamefont {Turok}},\ }\href {\doibase 10.1126/science.1070462} {\bibfield  {journal} {\bibinfo  {journal} {Science}\ }\textbf {\bibinfo {volume} {296}},\ \bibinfo {pages} {1436} (\bibinfo {year} {2002})},\ \Eprint {http://arxiv.org/abs/hep-th/0111030} {arXiv:hep-th/0111030} \BibitemShut {NoStop}%
\bibitem [{\citenamefont {van~de Heisteeg}\ \emph {et~al.}(2024{\natexlab{b}})\citenamefont {van~de Heisteeg}, \citenamefont {Vafa}, \citenamefont {Wiesner},\ and\ \citenamefont {Wu}}]{vandeHeisteeg:2022btw}%
  \BibitemOpen
  \bibfield  {author} {\bibinfo {author} {\bibfnamefont {D.}~\bibnamefont {van~de Heisteeg}}, \bibinfo {author} {\bibfnamefont {C.}~\bibnamefont {Vafa}}, \bibinfo {author} {\bibfnamefont {M.}~\bibnamefont {Wiesner}}, \ and\ \bibinfo {author} {\bibfnamefont {D.~H.}\ \bibnamefont {Wu}},\ }\href {\doibase 10.4310/bpam.2024.v1.n1.a1} {\bibfield  {journal} {\bibinfo  {journal} {Beijing J. Pure Appl. Math.}\ }\textbf {\bibinfo {volume} {1}},\ \bibinfo {pages} {1} (\bibinfo {year} {2024}{\natexlab{b}})},\ \Eprint {http://arxiv.org/abs/2212.06841} {arXiv:2212.06841 [hep-th]} \BibitemShut {NoStop}%
\bibitem [{\citenamefont {van~de Heisteeg}\ \emph {et~al.}(2023)\citenamefont {van~de Heisteeg}, \citenamefont {Vafa},\ and\ \citenamefont {Wiesner}}]{vandeHeisteeg:2023ubh}%
  \BibitemOpen
  \bibfield  {author} {\bibinfo {author} {\bibfnamefont {D.}~\bibnamefont {van~de Heisteeg}}, \bibinfo {author} {\bibfnamefont {C.}~\bibnamefont {Vafa}}, \ and\ \bibinfo {author} {\bibfnamefont {M.}~\bibnamefont {Wiesner}},\ }\href {\doibase 10.1002/prop.202300143} {\bibfield  {journal} {\bibinfo  {journal} {Fortsch. Phys.}\ }\textbf {\bibinfo {volume} {71}},\ \bibinfo {pages} {2300143} (\bibinfo {year} {2023})},\ \Eprint {http://arxiv.org/abs/2303.13580} {arXiv:2303.13580 [hep-th]} \BibitemShut {NoStop}%
\bibitem [{\citenamefont {Cribiori}\ and\ \citenamefont {L\"ust}(2023)}]{Cribiori:2023sch}%
  \BibitemOpen
  \bibfield  {author} {\bibinfo {author} {\bibfnamefont {N.}~\bibnamefont {Cribiori}}\ and\ \bibinfo {author} {\bibfnamefont {D.}~\bibnamefont {L\"ust}},\ }\href {\doibase 10.1002/prop.202300150} {\bibfield  {journal} {\bibinfo  {journal} {Fortsch. Phys.}\ }\textbf {\bibinfo {volume} {71}},\ \bibinfo {pages} {2300150} (\bibinfo {year} {2023})},\ \Eprint {http://arxiv.org/abs/2306.08673} {arXiv:2306.08673 [hep-th]} \BibitemShut {NoStop}%
\bibitem [{\citenamefont {Kachru}\ \emph {et~al.}(2003)\citenamefont {Kachru}, \citenamefont {Kallosh}, \citenamefont {Linde},\ and\ \citenamefont {Trivedi}}]{Kachru:2003aw}%
  \BibitemOpen
  \bibfield  {author} {\bibinfo {author} {\bibfnamefont {S.}~\bibnamefont {Kachru}}, \bibinfo {author} {\bibfnamefont {R.}~\bibnamefont {Kallosh}}, \bibinfo {author} {\bibfnamefont {A.~D.}\ \bibnamefont {Linde}}, \ and\ \bibinfo {author} {\bibfnamefont {S.~P.}\ \bibnamefont {Trivedi}},\ }\href {\doibase 10.1103/PhysRevD.68.046005} {\bibfield  {journal} {\bibinfo  {journal} {Phys. Rev. D}\ }\textbf {\bibinfo {volume} {68}},\ \bibinfo {pages} {046005} (\bibinfo {year} {2003})},\ \Eprint {http://arxiv.org/abs/hep-th/0301240} {arXiv:hep-th/0301240} \BibitemShut {NoStop}%
\bibitem [{\citenamefont {Townsend}\ and\ \citenamefont {Wohlfarth}(2004)}]{Townsend:2004zp}%
  \BibitemOpen
  \bibfield  {author} {\bibinfo {author} {\bibfnamefont {P.~K.}\ \bibnamefont {Townsend}}\ and\ \bibinfo {author} {\bibfnamefont {M.~N.~R.}\ \bibnamefont {Wohlfarth}},\ }\href {\doibase 10.1088/0264-9381/21/23/006} {\bibfield  {journal} {\bibinfo  {journal} {Class. Quant. Grav.}\ }\textbf {\bibinfo {volume} {21}},\ \bibinfo {pages} {5375} (\bibinfo {year} {2004})},\ \Eprint {http://arxiv.org/abs/hep-th/0404241} {arXiv:hep-th/0404241} \BibitemShut {NoStop}%
\bibitem [{\citenamefont {L\"ust}\ \emph {et~al.}(2019)\citenamefont {L\"ust}, \citenamefont {Palti},\ and\ \citenamefont {Vafa}}]{Lust:2019zwm}%
  \BibitemOpen
  \bibfield  {author} {\bibinfo {author} {\bibfnamefont {D.}~\bibnamefont {L\"ust}}, \bibinfo {author} {\bibfnamefont {E.}~\bibnamefont {Palti}}, \ and\ \bibinfo {author} {\bibfnamefont {C.}~\bibnamefont {Vafa}},\ }\href {\doibase 10.1016/j.physletb.2019.134867} {\bibfield  {journal} {\bibinfo  {journal} {Phys. Lett. B}\ }\textbf {\bibinfo {volume} {797}},\ \bibinfo {pages} {134867} (\bibinfo {year} {2019})},\ \Eprint {http://arxiv.org/abs/1906.05225} {arXiv:1906.05225 [hep-th]} \BibitemShut {NoStop}%
\bibitem [{\citenamefont {Li}\ \emph {et~al.}(2023)\citenamefont {Li}, \citenamefont {Palti},\ and\ \citenamefont {Petri}}]{Li:2023gtt}%
  \BibitemOpen
  \bibfield  {author} {\bibinfo {author} {\bibfnamefont {Y.}~\bibnamefont {Li}}, \bibinfo {author} {\bibfnamefont {E.}~\bibnamefont {Palti}}, \ and\ \bibinfo {author} {\bibfnamefont {N.}~\bibnamefont {Petri}},\ }\href {\doibase 10.1007/JHEP08(2023)210} {\bibfield  {journal} {\bibinfo  {journal} {JHEP}\ }\textbf {\bibinfo {volume} {08}},\ \bibinfo {pages} {210} (\bibinfo {year} {2023})},\ \Eprint {http://arxiv.org/abs/2306.02026} {arXiv:2306.02026 [hep-th]} \BibitemShut {NoStop}%
\bibitem [{\citenamefont {Palti}\ and\ \citenamefont {Petri}(2024)}]{Palti:2024voy}%
  \BibitemOpen
  \bibfield  {author} {\bibinfo {author} {\bibfnamefont {E.}~\bibnamefont {Palti}}\ and\ \bibinfo {author} {\bibfnamefont {N.}~\bibnamefont {Petri}},\ }\href {\doibase 10.1007/JHEP06(2024)019} {\bibfield  {journal} {\bibinfo  {journal} {JHEP}\ }\textbf {\bibinfo {volume} {06}},\ \bibinfo {pages} {019} (\bibinfo {year} {2024})},\ \Eprint {http://arxiv.org/abs/2405.01084} {arXiv:2405.01084 [hep-th]} \BibitemShut {NoStop}%
\bibitem [{\citenamefont {Shiu}\ \emph {et~al.}(2023{\natexlab{c}})\citenamefont {Shiu}, \citenamefont {Tonioni}, \citenamefont {Van~Hemelryck},\ and\ \citenamefont {Van~Riet}}]{Shiu:2022oti}%
  \BibitemOpen
  \bibfield  {author} {\bibinfo {author} {\bibfnamefont {G.}~\bibnamefont {Shiu}}, \bibinfo {author} {\bibfnamefont {F.}~\bibnamefont {Tonioni}}, \bibinfo {author} {\bibfnamefont {V.}~\bibnamefont {Van~Hemelryck}}, \ and\ \bibinfo {author} {\bibfnamefont {T.}~\bibnamefont {Van~Riet}},\ }\href {\doibase 10.1007/JHEP05(2023)077} {\bibfield  {journal} {\bibinfo  {journal} {JHEP}\ }\textbf {\bibinfo {volume} {05}},\ \bibinfo {pages} {077} (\bibinfo {year} {2023}{\natexlab{c}})},\ \Eprint {http://arxiv.org/abs/2212.06169} {arXiv:2212.06169 [hep-th]} \BibitemShut {NoStop}%
\bibitem [{\citenamefont {Shiu}\ \emph {et~al.}(2024{\natexlab{c}})\citenamefont {Shiu}, \citenamefont {Tonioni}, \citenamefont {Van~Hemelryck},\ and\ \citenamefont {Van~Riet}}]{Shiu:2023bay}%
  \BibitemOpen
  \bibfield  {author} {\bibinfo {author} {\bibfnamefont {G.}~\bibnamefont {Shiu}}, \bibinfo {author} {\bibfnamefont {F.}~\bibnamefont {Tonioni}}, \bibinfo {author} {\bibfnamefont {V.}~\bibnamefont {Van~Hemelryck}}, \ and\ \bibinfo {author} {\bibfnamefont {T.}~\bibnamefont {Van~Riet}},\ }\href {\doibase 10.1103/PhysRevD.109.066017} {\bibfield  {journal} {\bibinfo  {journal} {Phys. Rev. D}\ }\textbf {\bibinfo {volume} {109}},\ \bibinfo {pages} {066017} (\bibinfo {year} {2024}{\natexlab{c}})},\ \Eprint {http://arxiv.org/abs/2311.10828} {arXiv:2311.10828 [hep-th]} \BibitemShut {NoStop}%
\bibitem [{\citenamefont {Farakos}\ \emph {et~al.}(2023)\citenamefont {Farakos}, \citenamefont {Morittu},\ and\ \citenamefont {Tringas}}]{Farakos:2023nms}%
  \BibitemOpen
  \bibfield  {author} {\bibinfo {author} {\bibfnamefont {F.}~\bibnamefont {Farakos}}, \bibinfo {author} {\bibfnamefont {M.}~\bibnamefont {Morittu}}, \ and\ \bibinfo {author} {\bibfnamefont {G.}~\bibnamefont {Tringas}},\ }\href {\doibase 10.1007/JHEP10(2023)067} {\bibfield  {journal} {\bibinfo  {journal} {JHEP}\ }\textbf {\bibinfo {volume} {10}},\ \bibinfo {pages} {067} (\bibinfo {year} {2023})},\ \Eprint {http://arxiv.org/abs/2304.14372} {arXiv:2304.14372 [hep-th]} \BibitemShut {NoStop}%
\bibitem [{\citenamefont {Tringas}(2023)}]{Tringas:2023vzn}%
  \BibitemOpen
  \bibfield  {author} {\bibinfo {author} {\bibfnamefont {G.}~\bibnamefont {Tringas}},\ }\href@noop {} {\  (\bibinfo {year} {2023})},\ \Eprint {http://arxiv.org/abs/2309.16542} {arXiv:2309.16542 [hep-th]} \BibitemShut {NoStop}%
\bibitem [{\citenamefont {Demirtas}\ \emph {et~al.}(2020)\citenamefont {Demirtas}, \citenamefont {Kim}, \citenamefont {Mcallister},\ and\ \citenamefont {Moritz}}]{Demirtas:2019sip}%
  \BibitemOpen
  \bibfield  {author} {\bibinfo {author} {\bibfnamefont {M.}~\bibnamefont {Demirtas}}, \bibinfo {author} {\bibfnamefont {M.}~\bibnamefont {Kim}}, \bibinfo {author} {\bibfnamefont {L.}~\bibnamefont {Mcallister}}, \ and\ \bibinfo {author} {\bibfnamefont {J.}~\bibnamefont {Moritz}},\ }\href {\doibase 10.1103/PhysRevLett.124.211603} {\bibfield  {journal} {\bibinfo  {journal} {Phys. Rev. Lett.}\ }\textbf {\bibinfo {volume} {124}},\ \bibinfo {pages} {211603} (\bibinfo {year} {2020})},\ \Eprint {http://arxiv.org/abs/1912.10047} {arXiv:1912.10047 [hep-th]} \BibitemShut {NoStop}%
\bibitem [{\citenamefont {Demirtas}\ \emph {et~al.}(2021)\citenamefont {Demirtas}, \citenamefont {Kim}, \citenamefont {McAllister}, \citenamefont {Moritz},\ and\ \citenamefont {Rios-Tascon}}]{Demirtas:2021nlu}%
  \BibitemOpen
  \bibfield  {author} {\bibinfo {author} {\bibfnamefont {M.}~\bibnamefont {Demirtas}}, \bibinfo {author} {\bibfnamefont {M.}~\bibnamefont {Kim}}, \bibinfo {author} {\bibfnamefont {L.}~\bibnamefont {McAllister}}, \bibinfo {author} {\bibfnamefont {J.}~\bibnamefont {Moritz}}, \ and\ \bibinfo {author} {\bibfnamefont {A.}~\bibnamefont {Rios-Tascon}},\ }\href {\doibase 10.1007/JHEP12(2021)136} {\bibfield  {journal} {\bibinfo  {journal} {JHEP}\ }\textbf {\bibinfo {volume} {12}},\ \bibinfo {pages} {136} (\bibinfo {year} {2021})},\ \Eprint {http://arxiv.org/abs/2107.09064} {arXiv:2107.09064 [hep-th]} \BibitemShut {NoStop}%
\bibitem [{\citenamefont {Demirtas}\ \emph {et~al.}(2022)\citenamefont {Demirtas}, \citenamefont {Kim}, \citenamefont {McAllister}, \citenamefont {Moritz},\ and\ \citenamefont {Rios-Tascon}}]{Demirtas:2021ote}%
  \BibitemOpen
  \bibfield  {author} {\bibinfo {author} {\bibfnamefont {M.}~\bibnamefont {Demirtas}}, \bibinfo {author} {\bibfnamefont {M.}~\bibnamefont {Kim}}, \bibinfo {author} {\bibfnamefont {L.}~\bibnamefont {McAllister}}, \bibinfo {author} {\bibfnamefont {J.}~\bibnamefont {Moritz}}, \ and\ \bibinfo {author} {\bibfnamefont {A.}~\bibnamefont {Rios-Tascon}},\ }\href {\doibase 10.1103/PhysRevLett.128.011602} {\bibfield  {journal} {\bibinfo  {journal} {Phys. Rev. Lett.}\ }\textbf {\bibinfo {volume} {128}},\ \bibinfo {pages} {011602} (\bibinfo {year} {2022})},\ \Eprint {http://arxiv.org/abs/2107.09065} {arXiv:2107.09065 [hep-th]} \BibitemShut {NoStop}%
\end{thebibliography}%

\end{document}